\documentclass[prd,twocolumn,superscriptaddress,preprintnumbers,nofootinbib,floatfix,10pt]{revtex4}
\usepackage{epsfig}
\usepackage{color}
\usepackage{ulem}
\usepackage{amsmath}
\usepackage{amssymb}
\usepackage{subfigure}
\usepackage{array}
\usepackage{multirow}
\usepackage{booktabs}
\usepackage{natbib}
\definecolor{brown}{rgb}{0.93, 0.53, 0.18}

\def\nn{\nonumber}
\def\refe@jnl#1{{#1}}
\def\aj{\refe@jnl{Astron.~J.}}
\def\araa{\refe@jnl{Annu.~Rev.~Astron.~Astrophys.}}
\def\apj{\refe@jnl{Astrophys.~J.}}
\def\apjl{\refe@jnl{Astrophys.~J.~Lett.}}
\def\aap{\refe@jnl{Astron.~Astrophys.}}
\def\mnras{\refe@jnl{Mon.~Not.~R.~Astron.~Soc.}}
\def\prd{\refe@jnl{Phys.~Rev.~D}}
\def\fcp{\refe@jnl{Fund.~Cos.~Phys.}}
\def\physrep{\refe@jnl{Phys.~Rep.}}
\def\physlett{\refe@jnl{Phys.~Lett.}}

\def\invisible#1{  }

\newcommand{\be}{\begin{equation}}
\newcommand{\ee}{\end{equation}}
\newcommand{\beq}{\begin{equation}}
\newcommand{\eeq}{\end{equation}}
\newcommand{\bea}{\begin{eqnarray}}
\newcommand{\eea}{\end{eqnarray}}
\newcommand{\beaa}{\begin{eqnarray*}}
\newcommand{\eeaa}{\end{eqnarray*}}
\newcommand{\ba}{\begin{array}}
\newcommand{\ea}{\end{array}}
\newcommand{\bi}{\begin{itemize}}
\newcommand{\ei}{\end{itemize}}
\newcommand{\ben}{\begin{enumerate}}
\newcommand{\een}{\end{enumerate}}

\newcommand{\fdm}{ \psi_{DM} }
\newcommand{\lbar}{\bar{L}}
\newcommand{\smn}{\sigma_{\mu\nu}}
\newcommand{\bmn}{F^{\mu\nu}_Y}
\newcommand{\wmn}{F^{\mu\nu}_L}

\begin{document}
\preprint{ULB-TH/15-12}

\title{Dark Matter Decay to a Photon and a Neutrino:  \\the Double Monochromatic Smoking Gun Scenario}
\author{Cha\"imae El Aisati}
\affiliation{Service de Physique Th\'eorique - Universit\'e Libre de Bruxelles, 
Boulevard du Triomphe, CP225, 1050 Brussels, Belgium}
\author{Michael Gustafsson}
\affiliation{Institute for theoretical Physics - Faculty of Physics, Georg-August University G\"ottingen, Friedrich-Hund-Platz 1, D-37077 G\"ottingen, Germany }
\email{celaisat@ulb.ac.be; \\michael.gustafsson@theorie.physik.uni-goettingen.de; \\thambye@ulb.ac.be; tscarna@ulb.ac.be}
\author{Thomas Hambye}
\affiliation{Service de Physique Th\'eorique - Universit\'e Libre de Bruxelles, 
Boulevard du Triomphe, CP225, 1050 Brussels, Belgium}
\author{Tiziana Scarna}
\affiliation{Service de Physique Th\'eorique - Universit\'e Libre de Bruxelles, 
Boulevard du Triomphe, CP225, 1050 Brussels, Belgium}

\date{October 15, 2015}

\begin{abstract}
In the energy range from few TeV to 25 TeV, upper bounds on the dark matter decay rate into high energy monochromatic neutrinos have recently become comparable to those on monochromatic gamma-ray lines. 
This implies clear possibilities of a future {\it double} ``smoking-gun'' evidence for the dark matter particle, from the observation of both a gamma  and a neutrino line at the same energy. In particular, we show that a scenario where both lines are induced from the same dark matter particle decay leads to correlations that can already be tested.
We study this ``double monochromatic'' scenario by considering the complete list of lowest dimensional effective operators that could induce such a decay. 
Furthermore, we argue that, on top of lines from decays into two-body final states, three-body final states can also be highly relevant. In addition to producing a distinct hard photon spectrum, three-body final states also produce a line-like feature in the neutrino spectrum that can be searched for by neutrino telescopes.
\end{abstract}

\maketitle
\section{Introduction}

Thanks to new data released by the IceCube collaboration \cite{Aartsen:2014muf}, it has recently been shown in Ref.\ \cite{Aisati:2015vma} that constraints on the flux of monochromatic neutrinos in the TeV to 25 TeV energy range have largely improved and are now comparable to those holding on monochromatic photon fluxes from decaying dark matter (DM) \cite{Ackermann:2015lka,Abramowski:2013ax,Gustafsson:2013gca}. Within this energy range, which interestingly allows for thermally produced DM candidates, this opens the way to the study of a \textit{``double smoking gun''} DM particle evidence -- that is, the observation of both a \mbox{$\gamma$-line}
and a $\nu$-line
of similar intensities and energies. 
Three general scenarios can be considered, distinguished by whether these two spectral lines are produced from different annihilations channels, from different decay channels or from the same decay into a $\nu+\gamma$ final state (which is only possible for a decay). Along the first two scenarios,  $\gamma$ and $\nu$ fluxes could largely differ in terms of energy and intensity. This is not the case for the latter scenario, where both lines are directly correlated. We study in detail this last scenario and show that its observational  discovery could be around the corner.

To this end, we will consider the full list of lowest dimensional operators that induce such $\nu+\gamma$  decay channels, and analyze their associated phenomenology. As will be seen, for some operators it is only the 2-body DM decays that are relevant, while for other operators, and at high DM masses, the 3-body decay channels are actually dominant. In the latter case the $\nu+\gamma$ channel becomes subleading but, still,  these 3-body processes turn out to give interesting sharp spectral features -- similar to the internal Bremsstrahlung (IB) type --- both in the photon and in the neutrino spectra. We stress that both these types of spectral features  could be simultaneously observed. Thus on top of neutrino-\textit{line} searches, such an IB signal can also be looked for by neutrino telescopes (as explicitly done in \cite{Aisati:2015vma}). Besides sharp features, the operators also lead to the emission of a low energy continuum of cosmic rays (CRs), which leads to constraints we will determine too.

\section{Double-monochromatic scenario and effective operators}
\label{Sec:EffOperators}

\begin{figure}[t]
\center{\includegraphics[width=0.99 \columnwidth]{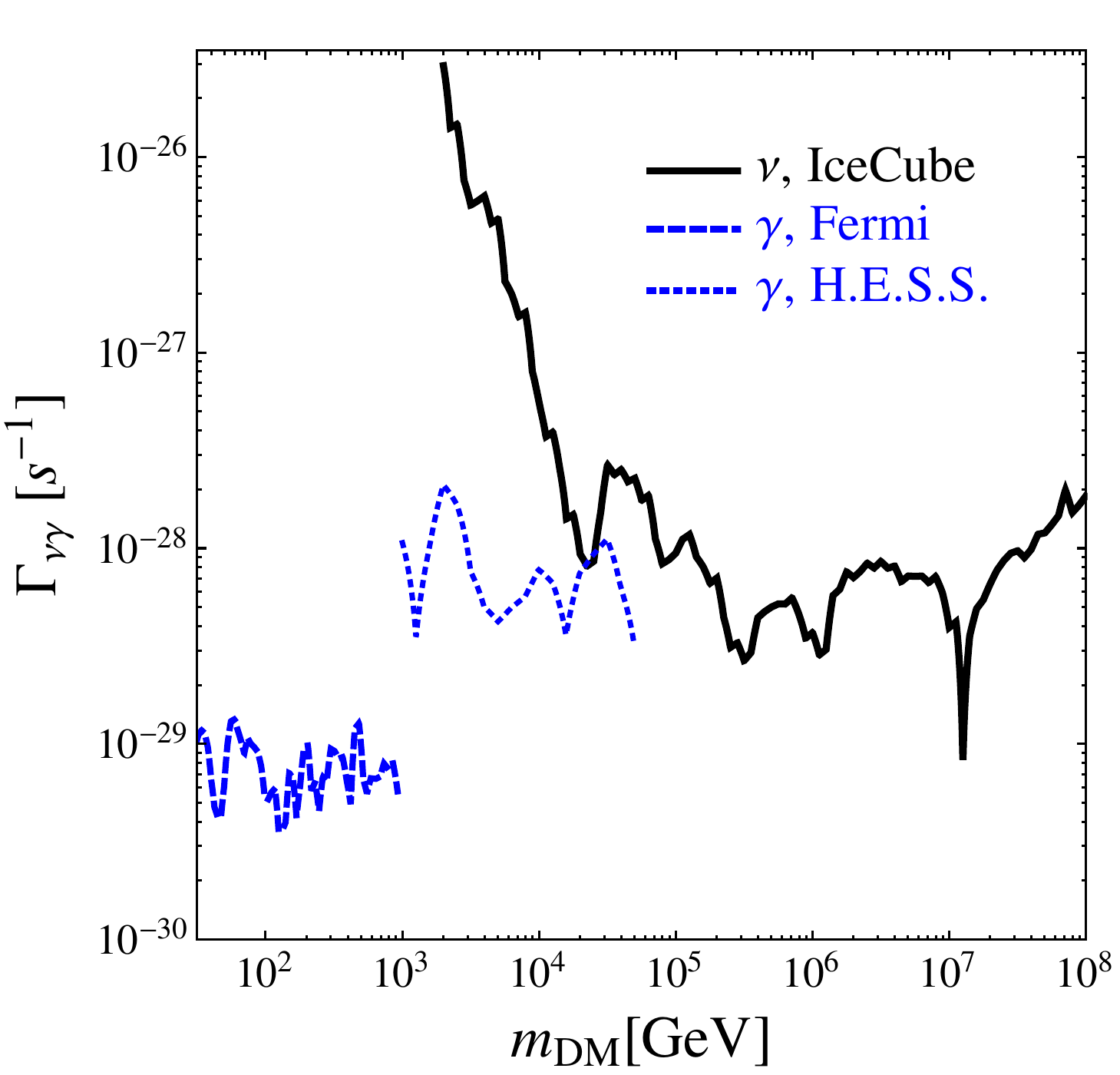}}
\caption{Current limits on $\Gamma(DM\rightarrow \gamma X)$, from Fermi-LAT \cite{Ackermann:2015lka} (dashed blue line) and H.E.S.S.\ \cite{Abramowski:2013ax,Gustafsson:2013gca} (dotted blue) and on  $\Gamma(DM\rightarrow \nu X)$, from Ref.~\cite{Aisati:2015vma} that used IceCube data (solid black). }
\label{fermiicecube}
\end{figure}

To begin with, let us first have a glance at Fig.~\ref{fermiicecube}, from Ref.~\cite{Aisati:2015vma}, which summarizes the current upper bounds on the $\Gamma_\gamma=\Gamma(DM\rightarrow \gamma +X)$ and $\Gamma_\nu=\Gamma(DM\rightarrow \nu +X)$ DM particle decay widths.

The figure shows that: i) For DM masses below few~TeV, constraints on gamma-line intensities are several orders of magnitudes stronger than those on neutrino lines; ii) Above the maximum energy considered by the H.E.S.S. collaboration, \textit{i.e.}\ $E_\gamma =25$~TeV, there are no numerically precise $\gamma$-line constraints (see however, e.g., \cite{Atkins:2004jf, Abdo:2009ku, Borione:1997fy, Schatz:2003aw, Abeysekara:2014ffg, Murase:2015gea,Esmaili:2015xpa}) while strong neutrino-line constraints exist up to energies several orders of magnitude beyond; iii) In the DM mass range from few TeV to 50 TeV, the constraints in the two signal channels are comparable and only differ by a factor of 1 to 5 in the range from 10 to 50~TeV and by a factor of 5 to 100 in the 2~TeV to 10~TeV DM mass range. Note also that sensitivities in both gamma- and neutrino-line searches are expected to improve within this (and an extended) energy range in the near future (by, e.g., CTA \cite{CTA} and IceCube \cite{Aartsen:2014oha,Aartsen:2014njl}).

A double ``smoking-gun'' evidence of a DM particle in the form of a $\gamma$- {\it and} a $\nu$-line appearing at the same energy could in principle happen for $\Gamma_{\nu}\gg\Gamma_\gamma$, $\Gamma_{\nu}\ll\Gamma_\gamma$ as well as for $\Gamma_{\nu}\sim\Gamma_\gamma$.
Nonetheless, the case in which the decay rates $\Gamma_{\nu}$ and $\Gamma_\gamma$ are similar is of special interest because such a correlation could indicate that the two lines are induced in the same process and thus potentially points towards the existence of decaying DM fermions.

As mentioned in the introduction, one can separate three main scenarios for the production of both  $\gamma$ and $\nu$ lines, namely production from different annihilation channels, from different decay channels or from a single decay into $\nu+\gamma$ (we do not consider more baroque mixtures of these scenarios). 

As is well known, annihilation scenarios can be distinguished from decay scenarios by looking at the sky morphology of the signal; the signal coming from an annihilation is quadratic in the DM density $\rho(r)$ -- and therefore more peaked towards the GC --- as opposed to a decay signal whose dependence is only linear on $\rho(r)$.\footnote{See, e.g., \cite{Ibarra:2013cra} and references therein.} 

A further possibility of distinguishing the scenarios then arises by looking at the relative energies and intensities of the lines.\footnote{Note that, in contrast to DM {\it decay} limits, current sensitivities on DM {\it annihilating rates} into monochromatic neutrinos \cite{Abbasi:2011eq,Aartsen:2014hva,Aartsen:2015xej} are much weaker than those into monochromatic gamma-rays \cite{Ackermann:2015lka,Abramowski:2013ax}; e.g.\ they differ by more than three orders of magnitude at a DM mass of 10~TeV.}
A fermion DM particle decay into $\nu+\gamma$ automatically predicts equal energy of the $\nu$ and $\gamma$ spectral lines. For the other two scenarios, where $\gamma$ and $\nu$ are produced through different annihilation or decay channels, the energies of the monochromatic $\nu$  and $\gamma$ are identical for $\nu\nu$ and $\gamma\gamma$ final states. In general however, when  $\nu A$ and $\gamma B$ final states are produced, neutrino and photon line energies will differ unless $A$'s and $B$'s masses are equal or negligibly small.  As for the relative intensity of the two lines, the decay channel $\psi_{DM} \rightarrow \nu \gamma$  stands out by its high predictivity. In this case, the two line intensities are clearly correlated, whereas for independent  production channels the ratio of these intensities could largely vary, depending on the explicit model considered.\footnote{Note that there is always a minimum degree of correlation: as a decay to $\nu\nu$ induces a decay to $\gamma\gamma$ at the two-loop level (or at the one-loop level if there is an associated $l^+l^-$ channel, an annihilation or decay into one channel implies the other process, but at a suppressed level.} 
In short, a clear possibility of a line signal in both gamma and neutrino telescopes, with similar intensities, arises if the DM particle is a fermion $\psi_{DM}$ that slowly decays into a $\gamma$ plus a $\nu$.\footnote{Other scenarios where a gamma and a neutrino line-like signals appear with a similar intensity and same energy are still possible. For example, in  multicomponent DM annihilation scenarios two distinct DM particles do not need to form a bosonic state (as annihilating conjugated particles) and could thus lead to a $\gamma+\nu$ annihilation final state. Double line-like signals might also arise from 3-body final states, such as from $\nu\bar\nu\gamma$  annihilation or decay processes. We will not consider these more elaborate cases.}

\medskip
To investigate what kind of monochromatic photon- to neutrino-flux ratios can be expected along this $\psi_{DM}  \rightarrow \nu\gamma$ scenario, we will use an effective theory approach. The motivation for using such an approach is clear.  The cosmological timescales required for the DM lifetime can naturally be explained in the framework of an accidental low energy global symmetry that is broken by new ultra-violet (UV) physics --- analogously to the proton decay case in the Standard Model (SM).
If the new scale of the UV physics $\Lambda$ is much higher than the electroweak scale and the DM particle mass $m_{DM}$, the decay rate is naturally low because it is suppressed by powers of $\Lambda$. In this case, the range of decay possibilities can be fully parameterized by determining the complete low-energy effective theory, order by order in inverse powers of $\Lambda$. It is known that, for a DM candidate with a mass roughly around the electroweak scale, a dimension 6 operator suppressed by 2 powers of the Grand Unified Theory scale $\Lambda_{GUT}$ leads to a DM particle lifetime around a value that is probed observationally today \cite{GUTdecay}.

Up to dimension 6, and assuming that the DM particle has spin 0, 1/2 or 1, it was shown in Ref.~\cite{Gustafsson:2013gca} that there exists only a quite limited list of operator structures leading to $\gamma$-lines.\footnote{See Ref.~\cite{Feldstein:2013kka} for a list of operators leading to monochromatic neutrinos.}
Disregarding, as in 
that  reference, the possibility of higher-spin DM particles (e.g.\ the gravitino that might decay into $\gamma+\nu$ \cite{Takayama:2000uz}), only eight of those operator structures also lead to a neutrino line via the $\gamma + \nu$ decay channel. Two of them are of dimension 5 
\begin{eqnarray}
{\cal O}^{(5)Y}&\equiv& \bar{L} \sigma_{\mu\nu} \psi_{DM} F^{\mu \nu}_Y,  \,\,\, \quad \psi_{DM}=(2,-1)\quad \,\,\,
\label{OpsiDM5Y}\\
{\cal O}^{(5)L}&\equiv& \bar{L} \sigma_{\mu\nu} \psi_{DM} F^{\mu \nu}_L,   \,\,\, \quad \psi_{DM}=(2/4,-1)
\label{OpsiDM5L}
\end{eqnarray}
and 6 are of dimension 6
\begin{eqnarray}
{\cal O}^{1Y}&\equiv& \bar{L} \sigma_{\mu\nu} \psi_{DM} F^{\mu \nu}_Y \phi,  \,\,\,  \, \,\, \psi_{DM}  \cdot \phi=(2,-1)\,\,
\label{Opsi1DMY}\\
{\cal O}^{1L}&\equiv& \bar{L} \sigma_{\mu\nu} \psi_{DM} F^{\mu \nu}_L \phi,    \,\,\,  \,  \,\, \psi_{DM} \cdot \phi=(2\oplus4,-1)\,\,
\label{Opsi1DML}\\
{\cal O}^{2Y}&\equiv& D_\mu\bar{L} \gamma_\nu  \psi_{DM} F^{\mu \nu}_Y,     \,\,\,    \,\,  \psi_{DM}=(2,-1) \quad\,\,\,\,
\label{Opsi2DMY}\\
{\cal O}^{2L}&\equiv& D_\mu\bar{L} \gamma_\nu  \psi_{DM} F^{\mu \nu}_L,     \,\,\,     \,\,  \psi_{DM}=(2/4,-1)\quad\,\,\,
\label{Opsi2DML}\\
{\cal O}^{3Y}&\equiv& \bar{L} \gamma_\mu D_\nu \psi_{DM} F^{\mu \nu}_Y,     \,\,\,     \,\, \psi_{DM}=(2,-1)\quad\,\,\,\,
\label{Opsi3DMY}\\
{\cal O}^{3L}&\equiv& \bar{L} \gamma_\mu D_\nu \psi_{DM} F^{\mu \nu}_L,     \,\,\,     \,\, \psi_{DM}=(2/4,-1)\quad\,\,
\label{Opsi3DML}
\end{eqnarray}
In the above list, $L$ represents a lepton doublet $L\equiv (\nu_L,l^-_L)^T$ of $e$, $\mu$ or $\tau$ flavor and $F^{\mu\nu}_{Y,L}$ represent the field strength tensors of the $U(1)_Y$  and $SU(2)_L$ gauge fields. 
The $(n,Y)$ labels, in the second column, denote the dimension $n$ and hypercharge $Y$ that the given field (or field combination) must have under the $SU(2)_L \times U(1)_Y$ group in order to guarantee gauge invariance. 
Whenever we refer to the contribution of an operator, we always mean the contribution of this operator plus the contribution of its hermitian conjugate.\footnote{  
That is, on top of the $\psi_{DM}$ decay channels, the hermitian conjugated operator induces $\bar{\psi}_{DM}$ decay to conjugated final states (with identical BRs).}
We will also assume the operators are fully flavor democratic (but results are only marginally affected by any other assumed flavor composition).

Beside the DM field, all operator structures above only involve SM fields, except the two operator structures of Eqs.~(\ref{Opsi1DMY}-\ref{Opsi1DML}). These two latter operator structures involve a scalar field $\phi$, which does not necessarily have to be the SM scalar doublet field, $H$. If we consider only SM fields in the operator and take into account all the various possibilities of DM multiplets and a complete set of $SU(2)_L$ index contractions, the operator structures above lead to 25 different effective operators. These 25 operators are listed in Table~\ref{tab:opp}.
\setlength{\extrarowheight}{3 pt}
\begin{table}[t]
\setlength{\extrarowheight}{3 pt}
{
\begin{tabular}{l|c|l|@{\hspace{0.5em}}l@{\hspace{0.5em}}}
\toprule
\textit{Operator} 								& DM field  		&  Fields contract. 	& \textit{Operator}  \\ 
\textit{Structure} 								& (\textit{n-plet}, $Y$)& ({\it n-plet})	& \textit{} 		 \\ 
\toprule 
$\lbar \smn \fdm \bmn$ 							& $(2, -1)$ 		& 					& ${\cal O}^{(5)Y}_\text{{2-let}}$	\\ 		\hline
\multirow{2}{*}{$\lbar \smn \fdm \wmn$} 				& $(2, -1)$  		& 					& ${\cal O}^{(5)L}_\text{{2-let}}$	\\   
 											& $(4, -1)$ 		& 					& ${\cal O}^{(5)L}_\text{{4-let}}$	\\ 		\hline
\multirow{2}{*}{$\lbar \smn \fdm \bmn H$}				& $(1, 0)$ 			& 					& ${\cal O}^{1Y}_{H,\text{{1-let}}}$	\\
 										 	& $(3, 0)$ 			& 					& ${\cal O}^{1Y}_{H,\text{{3-let}}}$	\\ 		\hline
\multirow{7}{*}{$\lbar \smn \fdm \wmn H$}			 	& $(1, 0)$ 			& 					& ${\cal O}^{1L}_{H,\text{{1-let}}}$ 	\\ 		\cline{2-4}
											& $(3, 0)$ 			& a: $(\lbar H) =1 $  		& ${\cal O}^{1L,a}_{H,\text{{3-let}}}$	\\
   											& $(3, 0)$   		& c: $(\fdm H) = 2 $  		& ${\cal O}^{1L,c}_{H,\text{{3-let}}}$	\\
											& $(3, 0)$  		& d: $(\fdm H) = 4 $     		& ${\cal O}^{1L,d}_{H,\text{{3-let}}}$	\\
 											& $(3, 0)$ 			& e: $(\lbar \fdm) =2 $ 		& ${\cal O}^{1L,e}_{H,\text{{3-let}}}$	\\
											& $(3, 0)$  		& f: $(\lbar \fdm) =4$   	& ${\cal O}^{1L,f}_{H,\text{{3-let}}}$	\\
 \cline{2-4}
 											& $(5, 0)$ 	 		& 					& ${\cal O}^{1L}_{H,\text{{5-let}})}$	\\  
\hline
$\lbar \smn \fdm \bmn  \tilde{H}$ 					& $(3, -2)$ 		&					& ${\cal O}^{1Y}_{\tilde{H},\text{{3-let}}}$	\\ 
\hline
\multirow{7}{*}{$ \lbar \smn \fdm \wmn \tilde{H}$ } 		& $(3, -2)$  		& b: $(\lbar \tilde{H}) = 3$ & ${\cal O}^{1L,b}_{\tilde{H},\text{{3-let}}}$	\\
  											& $(3, -2)$ 		& c: $(\fdm \tilde{H})= 2$  & ${\cal O}^{1L,c}_{\tilde{H},\text{{3-let}}}$	\\
  											& $(3, -2)$  		& d: $(\fdm \tilde{H}) = 4$  & ${\cal O}^{1L,d}_{\tilde{H},\text{{3-let}}}$\\
  											& $(3, -2)$  	 	& e: $(\lbar \fdm) = 2 $		& ${\cal O}^{1L,e}_{\tilde{H},\text{{3-let}}}$	\\
  											& $(3, -2)$     		& f: $(\lbar \fdm) = 4  $ 	& ${\cal O}^{1L,f}_{\tilde{H},\text{{3-let}}}$	\\ 		\cline{2-4}
 											& $(5, -2)$ 		&  					&${\cal O}^{1L}_{\tilde{H},\text{{5-let}}}$	 \\ 		\hline
 $ D_\mu \lbar \gamma_\nu \fdm \bmn $ 				& $(2, -1)$ 		& 					&${\cal O}^{2Y}_\text{{2-let}}$	 \\ 
 \hline
 \multirow{2}{*}{$ D_\mu\lbar \gamma_\nu  \fdm \wmn$} 	& $(2, -1)$ 		&					& ${\cal O}^{2L}_\text{{2-let}}$	\\ 
 											& $(4, -1)$  		&					& ${\cal O}^{2L}_\text{{4-let}}$\\ 		\hline
$\lbar \gamma_\mu  D_\nu \fdm  \bmn$ 				& $(2, -1)$ 		& 					& ${\cal O}^{3Y}_\text{{2-let}}$\\ 		\hline
\multirow{2}{*}{$\lbar \gamma_\mu D_\nu \fdm \wmn$} 	& $(2, -1)$ 		& 					&  ${\cal O}^{3L}_\text{{2-let}}$\\ 
 											& $(4, -1)$ 		& 					&  ${\cal O}^{3L}_\text{{4-let}}$\\ 
 \bottomrule
\end{tabular}
}
\caption{The ten possible effective operator {\it structures}, involving only SM fields beside the DM particle, for $DM\rightarrow\gamma\nu$ decay (1$^{st}$ column) with their allowed DM multiplets (2$^{nd}$ column) and various $SU(2)_L$ index contraction possibilities --- when not unique ---  of the fields in the operator (3$^{rd}$ column). The last column labels the 25 resulting effective operators (with the DM  multiplet, contraction choice and included scalar field specified in the label's indexes). 
 }
\label{tab:opp}
\end{table}

There are 3 cases from the dimension 5 operators in Eqs.~(\ref{OpsiDM5Y}-\ref{OpsiDM5L}) and 6 cases from the dimension 6 operators in Eqs.~(\ref{Opsi2DMY}-\ref{Opsi3DML}).
In addition, the operators involving a scalar field in Eqs.~(\ref{Opsi1DMY}-\ref{Opsi1DML}) lead to 9 cases when $\phi=H$ (where $\psi_{DM}$ is hyperchargeless)
\begin{eqnarray}
{\cal O}^{1Y}_H&\equiv& \bar{L} \sigma_{\mu\nu} \psi_{DM} F^{\mu \nu}_Y H,    \, \psi_{DM}  =(1/3,0)\quad\quad
\label{Opsi1DMYH}\\
{\cal O}^{1L}_H&\equiv& \bar{L} \sigma_{\mu\nu} \psi_{DM} F^{\mu \nu}_L H,     \, \psi_{DM} =(1/3_{a,\not{\hspace{0.3mm}b},c,d,e,f}/5,0)\quad\quad
\label{Opsi1DMLH}
\end{eqnarray}
and to 7 cases when $\phi= \tilde{H} \equiv i \sigma_2 H^*$ (where $\psi_{DM}$ has $Y=-2$) 
\begin{eqnarray}
{\cal O}^{1Y}_{\tilde{H}}&\equiv& \bar{L} \sigma_{\mu\nu} \psi_{DM} F^{\mu \nu}_Y \tilde{H},    \, \psi_{DM}  =(3,-2)\quad\quad
\label{Opsi1DMYHbar}\\
{\cal O}^{1L}_{\tilde{H}}&\equiv& \bar{L} \sigma_{\mu\nu} \psi_{DM} F^{\mu \nu}_L \tilde{H},     \, \psi_{DM} =(3_{\not{a},b,c,d,e,f}/5,-2).\quad\quad
\label{Opsi1DMLHbar}
\end{eqnarray}
Here, $H$ is the SM scalar doublet with hypercharge $-1$, i.e.\ $H=(H^0, H^-)$ with $H^0=(v + h+ia_0)/\sqrt{2}$, $v=174$ GeV and $m_h=125$~GeV. As indicated by the subscripts $\{a,b,c,d,e,f\}$, for both ${\cal O}^{1L}_{H}$ and ${\cal O}^{1L}_{\tilde{H}}$ there are various possible operators when $\psi_{DM}$ is a triplet because various contractions between the  $SU(2)_L$ indices of the fields are possible.
The six operator setups $3_{a,b,c,d,e,f}$ correspond to the case where $H$ and $L$ form a singlet or a triplet, where $\psi_{DM}$ and $H$ form a doublet or a quadruplet and where the $\psi_{DM}$ and $L$ form a doublet or a quadruplet (and correspondingly for the two remaining fields in the operator), respectively.\footnote{Only two out of these six contraction possibilities are linearly independent (they can all be written as linear combinations of the invariants obtained with, e.g., $\psi_{DM}\!\cdot\!\phi$ being a 2-let and a 4-let), but we study all these setups because they could in principle be induced by the mediation of a heavy multiplet with the corresponding quantum numbers.}  Note that for ${\cal O}^{1L}_{H,\text{3-let}}$  (${\cal O}^{1L}_{\tilde{H},\text{3-let}}$) the b (a) case does not lead to a DM decay into $\gamma \nu$ and therefore has to be excluded from the list, which is indicated by crossing out the $b$ and $a$ subscripts in Eqs.~\eqref{Opsi1DMLH} and \eqref{Opsi1DMLHbar}, respectively.

In the effective operator language, gauge invariance is manifest. This implies that any of the listed effective operators  necessarily gives a decay into e.g.\ $\nu Z$ in addition to the decay into $\nu \gamma$. This $Z$ channel does not only produce additional monochromatic neutrinos, but also leads to a continuum of CRs from subsequent $Z$ decays.

If $m_{DM}< m_Z$, the $\gamma \nu$ channel is the only line signal that is kinematically allowed  and, as a result, $\nu$ and $\gamma$ lines are both at the energy $m_{DM}/2$ with a relative intensity 
\begin{equation}
R_{\nu/\gamma}\equiv \frac{n_\nu}{n_\gamma}=1,
\end{equation}
where $n_i$ refers to the number of particles of type ``i'' produced per DM particle decay.

If instead $m_{DM}>m_Z$, the $\gamma \nu$ and $\nu Z$ channels give one $\gamma$ line and two $\nu$-lines. If both neutrino peaks are resolved, one is at $E_\nu = m_{DM}/2$ with the same intensity as the $\gamma$-line and the other is at the energy
\begin{equation}
E_\nu = \frac{m_{DM}}{2} \left(1- \frac{m_Z^2}{m_{DM}^2} \right).
\end{equation} 
In this case, the intensity ratio of the lower energy neutrino line (from $\nu Z$) to the gamma line (from $\nu \gamma$)  is equal to \begin{equation}
 R_{\nu/\gamma} = \tan^2 \theta_W \simeq 0.3 \;\; \mathrm{or} \;\; \tan^{-2} \theta_W \simeq 3.3,
\end{equation} 
depending on whether the effective operator involves a $F_Y^{\mu\nu}$  or a $F_L^{\mu\nu}$ field strength, respectively.  

When $m_{DM} \gg m_Z$, the two neutrino peaks are very close to each other in energy and, given the finite experimental resolutions, they are inseparable. Therefore, they can be summed into one single effective neutrino line at energy $\sim m_{DM}/2$ that is more intense than the $\gamma$-line. For instance, for a $10\%$ energy resolution this is typically a good approximation when $m_{DM} \gtrsim 300$~GeV. In this case, for the operators that contain a hypercharge field strength $F^{\mu\nu}_Y$, the ratio of line intensities is
\begin{equation}
\label{Ynugammaratioratio}
R_{\nu/\gamma} =\frac{1}{\cos^2 \theta_W}\simeq1.3\,,
\end{equation}
whereas, for operators involving the SU(2)$_L$ field strength $F^{\mu\nu}_L$, this ratio is larger
\begin{equation}
\label{Lnugammaratioratio}
R_{\nu/\gamma} =\frac{1}{\sin^2 \theta_W}\simeq 4.3\,.
\end{equation}

For high DM masses,  all of our operators necessarily  give one of these two latter predictions for the DM decays into 2-body final states.
Hence, the field strength that is contained in the active operator can in principle be experimentally distinguished.
In practice, nevertheless, this might not always be so simple. If several operators are induced by the UV physics, all will contribute and, in some cases, interfere. In fact, a contribution from several operators is to be expected in many cases but is not mandatory.
Models with operators involving only one type of field strength up to dimension 6 can easily be found, see the example of section \ref{Sec:quintuplet} below. Unless particular destructive interferences among several operators with both types of field strengths take place, a ratio measured of order of a few would constitute a strong indication for the single $\gamma\nu$ decay channel scenario (but not a proof), whereas larger values would constitute a strong indication for separate channel scenarios. Monochromatic line intensity ratios $R_{\nu/\gamma}$ smaller than 1 would definitely exclude this $\gamma\nu$ channel scenario and require a separate channel scenario --- since for each $\gamma$ at least one $\nu$ is produced.\footnote{Note that the values of $R_{\nu/\gamma}$ stated here hold for DM decays into 2-body final states, i.e.\ they apply to the operators that do not involve any relevant 3-body decay channel. See Fig.~\ref{summaryratios} below for the operators with a scalar field, Eqs.~(\ref{Opsi1DMYH})-(\ref{Opsi1DMLHbar}), where 3-body decays into line-like signals matter.}

\section{Associated cosmic ray fluxes}
\label{Sec:associatedCR}
Besides producing monochromatic fluxes of photons and neutrinos, the operators above lead to the emission of a continuum of CRs from $\nu Z$ and, in some cases, $W^\pm l^\mp_L$ decay channel(s). For the dimension 5 and 6 operators above, this has already been analysed at length in Ref.~\cite{Gustafsson:2013gca}.
As explained (and defined) in Ref.~\cite{Gustafsson:2013gca}, each operator leads to a well defined ratio between the number of emitted monochromatic $\gamma$ and the amount of CR, $R_{\gamma/CR}\equiv n_\gamma/n_{CR}$. The 25 operators above turn out to lead to only 5 possible ratios, that we call A, C, D, E and F.

Prediction A refers to the ratio
\be
A: \quad R_{\gamma/CR} =\cos^2\theta_W/(\sin^2\theta_W \cdot n_{CR/Z}), 
\label{Aratio}
\ee
where $n_{CR/Z}$ is the number of CRs of a given particle type (e.g.\ antiprotons) produced per $Z$ decay, with $\theta_W$ the Weinberg angle. This A ratio is the largest possible ratio, i.e.\ the lowest amount of associated CRs one can have from the full list of operators.
Prediction C refers to 
\be
C: \quad R_{\gamma/CR}=\sin^2\theta_W/(\cos^2\theta_W \cdot n_{CR/Z}). 
\ee
As for the D, E and F ratios they apply to operators which lead to decay channels involving the $W$, \mbox{$\psi_\mathrm{DM} \rightarrow W^\pm l^\mp $},
\bea
&&\!\!D,E,F: \quad R_{\gamma/CR}= \nn\\
&&\!\!
\frac{\sin^2\theta_W}
{\cos^2\theta_W \cdot n_{CR/Z}+c_W \cdot (n_{CR/ W^+ l^- }+n_{CR/ W^- l^+ })},\quad\quad
\label{DEratios}
\eea
with $c_W=1/4, \,1, \,9/4$ for the $D,\,E$ and $F$ ratio respectively. 
\begin{table}
\aboverulesep = 0.2ex
\belowrulesep = 0.2ex

\centering
\begin{tabular*}{221pt}  {  
@{\hspace{0.2em}}c@{\hspace{0.7em}}   
@{\hspace{0.2em}}c@{\hspace{0.9em}}   |
@{\hspace{0.5em}}l@{\hspace{1.5em}}  |
@{\hspace{0.0em}}c@{\hspace{1.1em}} | 
@{\hspace{0.6em}}c@{\hspace{0.6em}}  }
\toprule

		& \hspace{-1.3cm}{\it DM field}	   			& $Operator$								&   \multicolumn{2}{c}{ \hspace{-0.3cm} \it Prediction} \\
{\it n-plet},	& $Y$\;        							&										&\hspace{0.3cm}$R_{\nu/\gamma}$		 	& $R_{\gamma/CR}$  \\
\midrule[0.3ex]
\multirow{2}{*}{1}			& \multirow{2}{*}{0}		& ${\cal O}^{1Y}_H$							&\;\;  1.3  			& A    			\\ \cline{4-5} 
						&					&${\cal O}^{1L}_{H}$  						&\;\;  4.3			& E				\\
\midrule[0.2ex]
\multirow{2}{*}{2}			& \multirow{2}{*}{-1}		& ${\cal O}^{(5)Y}$, ${\cal O}^{2Y}$, ${\cal O}^{3Y}$	&\;\;  1.3  			& A    			 \\ \cline{4-5}
						&					&${\cal O}^{(5)L}$, ${\cal O}^{2L}$, ${\cal O}^{3L}$  	&\;\;  4.3			& E				\\
\midrule[0.2ex]
\multirow{4}{*}{3}			& \multirow{4}{*}{0}		& ${\cal O}^{1Y}_H$							&\;\;  1.3  			& A    			 \\
						&					& ${\cal O}^{1L,a}_{H}$  						&\;\;  4.3			& C				\\
						&					& ${\cal O}^{1L,d}_{H}$, ${\cal O}^{1L,f}_{H}$		&\;\;  4.3			& D				\\
						&					& ${\cal O}^{1L,c}_{H}$, ${\cal O}^{1L,e}_{H}$		&\;\;  4.3			& E				\\
\midrule[0.2ex]
\multirow{5}{*}{3}			& \multirow{5}{*}{-2}		&  ${\cal O}^{1Y}_{\tilde{H}}$					&\;\;  1.3  			& A    			 \\ \cline{4-5}
						&					&  ${\cal O}^{1L,e}_{\tilde{H}}$				&\;\;  4.3			& C				\\
						&					&  ${\cal O}^{1L,b}_{\tilde{H}}$, ${\cal O}^{1L,d}_{\tilde{H}}$	&\;\;  4.3			& D				\\
						&					&  ${\cal O}^{1L,c}_{\tilde{H}}$				&\;\;  4.3			& E				\\
						&					&  ${\cal O}^{1L,f}_{\tilde{H}}$				&\;\;  4.3			& F				\\
\midrule[0.2ex]
\multirow{1}{*}{4}			& \multirow{1}{*}{-1}		&  ${\cal O}^{(5)L}$, ${\cal O}^{2L}$, ${\cal O}^{3L}$	&\;\;  4.3  			& D    			 \\
\midrule[0.2ex]
\multirow{1}{*}{5}			& \multirow{1}{*}{0}		& ${\cal O}^{1L}_{H}$ 						&\;\;  4.3  			& D    			 \\
\midrule[0.2ex]
\multirow{1}{*}{5}			& \multirow{1}{*}{-2}		&  ${\cal O}^{1L}_{\tilde{H}}$ 					&\;\;  4.3  			& D    			 \\					
\bottomrule
\end{tabular*}
\caption{
Predicted phenomenology of the possible DM setups from all the effective operators that give $DM\rightarrow \gamma\nu$ decays. $R_{\nu/\gamma}$ gives the $\nu$- to $\gamma$-line intensity ratio and $R_{\gamma/CR}$  the amount of associated CRs as defined in Eqs.~(\ref{Aratio}-\ref{DEratios}). The operators are defined in Table~\ref{tab:opp} (omitting their ``DM n-plet'' index as it is set by the 1$^\mathrm{st}$ column). These are the predictions from DM decays into 2-body final states. For DM masses above $\sim 4$~TeV, the predictions from the operators including a $H$ or $\tilde{H}$ are modified by 3-body decays, and these are studied in Section~\ref{sec:3bdy}.}
\label{tab:pheno}
\end{table}

These A, C, D, E, F predictions hold when the DM dominantly decays into 2-body final states. Table \ref{tab:pheno} gives which one of the five  $R_{\gamma/CR}$ ratios each of the 25 operator setups gives. For each possible DM field representation, the table lists the possible effective operators together with their corresponding $R_{\nu/\gamma}$ line ratio (2 possible)  and  line to CR continuum $R_{\gamma/CR}$ ratio (5 possible)  predictions.
The A ratio is the largest ratio one can have and is obtained for all the operators which involve a $F_Y$ field strength. The four C, D, E and F ratios are obtained from the operators involving a $F_L$ field strength.\footnote{Note that if in Eqs.~(\ref{Opsi1DMY})-(\ref{Opsi1DML}), the $\phi$ field is not the SM scalar doublet but a BSM field, then there are many more possibilities depending on the quantum numbers of this field, and we will not consider them. However, it is worth to note that whatever the multiplet is, the operators of Eqs.~(\ref{Opsi1DMY}) and (\ref{Opsi1DML}) cannot give a $R_{\gamma/CR}$ ratio larger than the $A$ and $C$ ratio respectively.}

\begin{figure}[t]
\vspace{-0.6cm}
\includegraphics[width=1.08 \columnwidth]{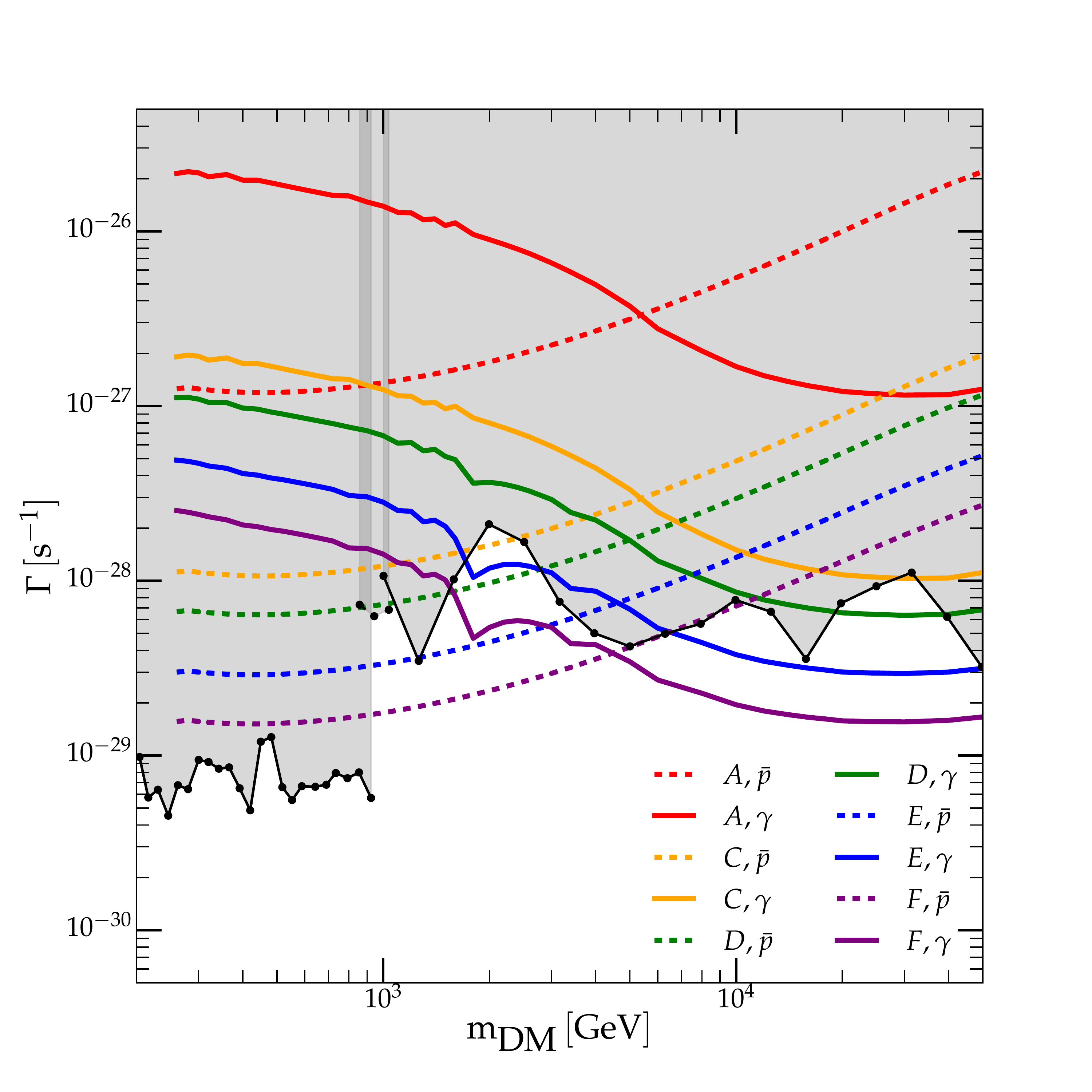}
\caption{$95\%$ CL limits on DM decay rates to monochromatic gamma-ray lines. Excluded regions from direct line searches \cite{Ackermann:2015lka, Abramowski:2013ax, Gustafsson:2013gca} (grey regions) as well as indirect upper bounds derived from the associated CR emission each operator unavoidably induces  (colored dotted curves are constraints from antiproton data \cite{Adriani:2010rc}, and solid curves from gamma-ray data \cite{Ackermann:2014usa}). Shown are the bounds we get for the A, C, D, E, F ratios given in Eqs.~(\ref{Aratio})-(\ref{DEratios}) when all 2-body DM decays are included. 
These cases apply to the various possibilities of operators and DM multiplets according to Table~\ref{tab:opp} and \ref{tab:pheno}, as explained in the text.
}
\label{CRIBbounds2body}
\end{figure}

The colored lines in Fig.~\ref{CRIBbounds2body} show the corresponding upper bounds on the decay width into monochromatic photons, $\Gamma(\psi_{DM} \rightarrow \nu \gamma)$, obtained by imposing that the associated fluxes of CRs (antiprotons and continuum photons) induced by each operator do not exceed the observational bound on these CR fluxes. The antiproton (continuum photon) constraints give the best limit for $m_{DM}$ below (above) $\sim 5$~TeV.
For the operators of Eqs.~(\ref{OpsiDM5Y})-(\ref{OpsiDM5L}) and (\ref{Opsi2DMY})-(\ref{Opsi3DML}), these bounds are valid up to the contribution of decays with 3 or more bodies in the final state. These extra 3-body contributions can, however, be safely neglected because the branching ratios (BRs) of these channels are phase-space suppressed. For the operators of (\ref{Opsi1DMY})-(\ref{Opsi1DML}) (and hence Eqs.~(\ref{Opsi1DMYH})-(\ref{Opsi1DMLHbar})) these bounds are also reliable, unless $m_{DM} \gg v_\phi$, with $v_\phi$ the vacuum expectation value of the $\phi$ field.  If $m_{DM} \gg v_\phi$, the 3-body contribution is no longer subleading, see Section IV below.

\medskip
The bounds of Fig.~\ref{CRIBbounds2body} have been obtained in the same way as in Ref.~\cite{Gustafsson:2013gca}, using the updated isotropic gamma-ray background measurement from the Fermi-LAT \cite{Ackermann:2014usa} (up to 820 GeV photon energies) and the H.E.S.S.\ telescope data \cite{Abramowski:2013ax} (up to 25 TeV). In practice, when considering only 2-body final states, and up to $m_{DM}\sim  50$~TeV at least, the constraints from Fermi-LAT are always stronger than those found from this H.E.S.S.\ data. 

We assumed a NFW profile for the DM density \cite{Navarro:1995iw}, with a local density of $\rho_{\odot}= 0.39$~GeV/cm$^{3}$ and a scale radius $r_s=24.42$~kpc.  For the CR calculation we used the ``PPPC'' code  \cite{Cirelli:2010xx}. For antiproton flux calculations, we used  the ``MED'' propagation model and assumed that conventional anti-proton background contributions exactly match the used PAMELA data \cite{Adriani:2010rc} in each energy bin. For the continuum photon fluxes,   the background flux is assumed to be a completely free (positive definite) function of the photon energy, i.e.\ the photon flux from DM  makes up all the flux in any bin where it overshoots the data. From these setups we derive our 95\;\%\;C.L.\ limits on any DM signal by a $\chi^2$ fit and requiring that $\chi^2 < 3.84$. We have included electroweak corrections when calculating the continuum energy spectrum of gamma-rays and antiprotons from DM particle decays. 
These electroweak corrections might also induce line-like signals and we will comment on this in section \ref{sec:EWline}, but they are not added to our predicted line-signals. For further comments on our astrophysical constraints, see Appendix~\ref{sec:A}.

\medskip
For the operators including ``$F_Y$'', i.e.~A cases, the direct limits on monochromatic gamma-rays are much stronger than the derived constraints from associated CRs;  by a factor of about 10 to 100 for $m_{DM}$ in the range of  1~TeV to 50~TeV.

For the operators including ``$F_L$'', i.e.~$C$ to $F$ cases, 
 the indirect CR bounds at the 2-body level are competitive with direct line searches (within a factor 10) as soon as $m_{DM}\gtrsim 1$~TeV (thanks to the updated continuum photon constraints, as opposed to the bounds reported in Ref.~\cite{Gustafsson:2013gca}).

Table \ref{tab:pheno} and Fig.~\ref{CRIBbounds2body}  thus summarize the results we have obtained so far, i.e.\ the expected  phenomenology for the operators giving DM decay into  2-body SM final states.  Clearly, some operators lead to the same phenomenology for indirect DM probes.

For example, all operator structures including the field strength $F_Y$ (i.e., ${\cal O}^{(5)Y}$, ${\cal O}^{2Y}$, ${\cal O}^{3Y}$, ${\cal O}^{2Y}_H$ and  ${\cal O}^{2Y}_{\tilde{H}}$) give the same $R_{\nu/\gamma} \simeq 1.3$ and have the same CR prediction A. 
At the same time, Fig.~\ref{CRIBbounds2body} shows that, for the operators with  $R_{\nu/\gamma} \simeq 4.3$, there are real further possibilities to distinguish between the operators because they predict the C to F cases in the amount of associated  CRs.
In the most minimal setups, DM particles with $Y \ne 0$ can also be strongly constrained by direct DM searches due to that $Z$ boson mediations induce direct interactions with nuclei.  Various mechanisms could however be invoked to avoid this, e.g.\  by mass splittings within an $SU(2)$ DM multiplet or if the DM field appropriately mixes with a pure SM singlet.  All operators can therefore be valid, but one has to keep in mind that a given DM field might also give a different signal in direct DM searches that could be used to differentiate the phenomenology of the different operators even further.

Depending on the UV completion, there might obviously also be new linear combinations when there are several possible operators for a given DM field.  However, in general it requires a careful tuning of possible interferences to significantly alter individual operators predictions \cite{Scarna:2014th}.

\section{Importance of 3-body decays for
operators involving a scalar field: line-like signals.}
\label{sec:3bdy}

For the operators involving a scalar field $\phi$, namely  Eqs.~(\ref{Opsi1DMY}) and (\ref{Opsi1DML}), there are, beside 2-body decays, 
3-body decay processes $\psi_{DM} \rightarrow \nu \gamma \phi$, $\psi_{DM} \rightarrow \nu Z \phi$, $\psi_{DM} \rightarrow \nu W \phi$ and $\psi_{DM} \rightarrow  l^- W^+ \phi$ (and decays to the CP conjugated states). 
While the BRs of the former ones are proportional to $v_\phi^2$, the BRs of the latter are proportional to $m^2_{DM}$. Therefore, the 3- to 2-body decay width ratio scales as $\sim m^2_{DM}/64 \pi^2v^2_\phi$. As a result, for these two operator structures, unlike for all other operators where 3-body final states are expected to only give a subleading contribution, the 3-body decay channel will dominate the decay width for large enough values of $m_{DM}/v_\phi$. 

If $\phi$ is the SM scalar doublet, as in {Eqs.~(\ref{Opsi1DMYH})-(\ref{Opsi1DMLHbar})},
then the 3-body decays can start to dominate for $m_{DM}\gtrsim 4$~TeV. 
In this case, replacing the Goldstone bosons by their corresponding longitudinal gauge bosons (in the  unitary gauge) the list of possible 3-body decays is
\begin{align*}
\psi_{DM}\rightarrow &\nu \gamma h, &\nu \gamma Z_L,  \qquad &  l \gamma W_L,\\
 &\nu Z h, &\nu Z Z_L,  \qquad  & l Z W_L,\\
 &l W h, & lWZ_L, \qquad & \nu W W_L .
\end{align*} 
Above 4~TeV it is a quite good approximation to calculate these 3-body decays in the electroweak unbroken phase, by calculating the 
$\psi_{DM}\rightarrow  \nu W_3 H^0,l W_3 H^+, \nu W H^+, l W H^0$ decay widths and then use the equivalence theorem to relate them to the broken phase associated processes. 

Note that different multiplets which give the same line to CR ratio at 2-body decay level do not necessarily have the same decay channels and BRs for the 3-body decays.
In the following, we will consider a representative set of examples which allows to see what is the typical range of possibilities. We will consider the following 4 cases with $\phi = H$, i.e.\  with the DM particle having $Y=0$,
\begin{align}
&\text{$\tilde{A}$} :	  &&{\cal O}^{1Y}_{H,\mathrm{1/3-let}}  &&\hspace{-0.5cm}\equiv  &&\bar{L} \sigma_{\mu\nu} \psi_{DM}^\mathrm{1/3-let} F^{\mu \nu}_Y H \,\,
\label{A3body}\\
&\text{$\tilde{C}$} :	  &&{\cal O}^{1L,a}_{H,\mathrm{3-let}} 	&&\hspace{-0.5cm}\equiv  &&\bar{L} \sigma_{\mu\nu} \psi_{DM}^\mathrm{3-let} F^{\mu \nu}_L H    \,\,\,\,\,\,\,\,\,
\label{C3body}\\
&\text{$\tilde{D}$} :	  &&{\cal O}^{1L}_{H,\mathrm{5-let}} 	&&\hspace{-0.5cm}\equiv  &&\bar{L} \sigma_{\mu\nu} \psi_{DM}^\mathrm{5-let} F^{\mu \nu}_L H    \,\,\,\,\,\,\,\,\,
\label{D3body}\\
&\text{$\tilde{E}$} :	  &&{\cal O}^{1L}_{H,\mathrm{1-let}}	&&\hspace{-0.5cm}\equiv  &&\bar{L} \sigma_{\mu\nu} \psi_{DM}^\mathrm{1-let} F^{\mu \nu}_L H    \,\,\,\,\,\,\,\,\,
\label{E3body}
\end{align}
In the first case, $\psi_{DM}$ has to be a singlet (1-let) or a triplet (3-let). Both give the same phenomenology.
In the subsequent cases, $\psi_{DM}$ 
is a triplet (3-let, with $\bar{L}$ and $H$ forming a singlet), quintuplet (5-let) and singlet (1-let), respectively.
As indicated, we will denote these cases $\tilde{A}$, $\tilde{C}$, $\tilde{D}$ and $\tilde{E}$ according to the ratios they give at the 2-body decay level, i.e.~Eqs.~(\ref{Aratio})-(\ref{DEratios}), but with an additional ``tilde" to stress that, for these cases, 3-body channels are dominant at high mass.

\begin{table}[t]
\centering
\begin{tabular}{c|c|ccc}
\toprule 
\\
\multirow{1}{*}{ $\begin{array}{c}Decay\\Channel\end{array} $}   
& \multicolumn{4}{c}{\textit{Operator}}  \\  
 & $F_Y$ &  \multicolumn{3}{c}{$F_L$} \\
&$\tilde{A}$ & $\tilde{E}$ &$\tilde{C}$ & $\tilde{D}$ \\
&(1/3-let) & (1-let) &(3-let) & (5-let) \\
\hline 
 & & & & \\
$\nu \gamma $  	& 4 $\cos^2\theta_W$   	& 4 $ \sin^2 \theta_W$  	& 4 $ \sin^2 \theta_W$ 	& 4 $ \sin^2 \theta_W$\\
$\nu  Z $  			& 4 $\sin^2\theta_W$   	& 4 $ \cos^2 \theta_W$  	& 4 $ \cos^2 \theta_W$ 	& 4 $ \cos^2 \theta_W$\\
$l W $  			& 0   					& 8 					& 0  					& 2\\
$\nu \gamma h $  	& $\cos^2\theta_W$   	& $ \sin^2 \theta_W$  	& $ \sin^2 \theta_W$ 	& $ \sin^2 \theta_W$\\
$\nu \gamma Z_L $  & $\cos^2 \theta_W $ 	& $ \sin^2 \theta_W$   	& $ \sin^2 \theta_W$  	& $ \sin^2 \theta_W$\\
$ l \gamma W_L $  	&  2 $\cos^2\theta_W$  	&2  $ \sin^2 \theta_W$  	& 2 $ \sin^2 \theta_W$ 	& 2 $ \sin^2 \theta_W$\\
$\nu Z h $  		& $ \sin^2 \theta_W$  	& $\cos^2\theta_W$    	& $\cos^2\theta_W$   	& $\cos^2\theta_W$\\
$\nu Z Z_L$ 		& $\sin^2 \theta_W$ 		& $\cos^2\theta_W$ 		& $\cos^2\theta_W$     	& $\cos^2\theta_W$ \\
$ l Z W_L$		& 2 $\sin^2 \theta_W$  	& 2 $\cos^2\theta_W$ 	& 2 $\cos^2\theta_W$   	&2 $\cos^2\theta_W$ \\
$  l W h $			&0					&2  					&0 					&1/2\\
$  l WZ_L$		&0 					&2  					&0 					& 1/2\\
$ \nu W W_L$		&0					&4  					&0 					&1 \\
 & & & & \\
\bottomrule
\end{tabular}
\caption{Branching ratios of the 3-body processes induced by the ``$F_Y$'' and ``$F_L$'' operators of Eqs.~(\ref{A3body})-(\ref{E3body}), up to the factors given in Eqs.~(\ref{eq:br2b}) and (\ref{eq:br3b}).}
\label{tab:BR}
\end{table}
The BRs of all decay channels for these 4 cases are given in Tab.~\ref{tab:BR}, up to the normalization factors given by
\begin{align}
\frac{c \; (64 \pi^2 v^2)}{m_{DM}^2+64 \pi^2 v^2} \quad \textnormal{for 2-body decays},
\label{eq:br2b}\\
\frac{c \; m_{DM}^2}{m_{DM}^2+64 \pi^2 v^2} \quad \textnormal{for 3-body decays},
\label{eq:br3b}
\end{align}
where $c$ is a constant equal to $1/4$, $1/4$, $1/6$, $1/12$ respectively. From Eq.~(\ref{eq:br3b}), it is clear that, due to the relative $m_{DM}^2/v^2$ factor, the various 2-body BRs which dominate for $m_{DM}$ around the electroweak scale get negligible at higher masses.

Out of these various channels, it is useful to define the primary $\gamma$ and primary $\nu$ decay widths, at 2- and 3-body levels, 
\begin{eqnarray}
\Gamma_\gamma^{2b}&\equiv&\Gamma_{\nu\gamma},\\
\Gamma_\nu^{2b}&\equiv&\Gamma_{\nu \gamma}+ \Gamma_{\nu Z},\\
\Gamma_\gamma^{3b}&\equiv& \Gamma_{\nu \gamma h}+\Gamma_{\nu \gamma Z_L}+\Gamma_{ l \gamma W_L}\,,\\ 
\Gamma^{3b}_{\nu}&\equiv&\Gamma_{\nu \gamma h} +\Gamma_{\nu \gamma Z_L} + \Gamma_{\nu Z_L h} \nonumber\\
&+& \Gamma_{\nu Z_L Z}+ \Gamma_{\nu W_L W}\,.
\end{eqnarray}

The effect of these 3-body decays is triple.  First, they bring a hard primary photon contribution of the IB type, normalized by $\Gamma^{3b}_{\gamma}$, that has an energy spectrum rapidly increasing 
\be
\frac{dN}{dE_\gamma} = \frac{64}{m_{DM}} \; \left( \frac{E_\gamma}{m_{DM}} \right)^3 \;  \Theta\left(\dfrac{m_{DM}}{2}-E_\gamma\right)
\label{IBgspectrum}
\ee
and which peaks in intensity at the kinematic cutoff $E_\gamma=m_{DM}/2$ (up to ${\cal{O}}(m_h^2/m_{DM}^2)$ corrections).\footnote{Both line and IB spectral features show up together in many frameworks  (e.g.\ \cite{susy, Ibarra:2014qma}), from tree-level 3-body radiative annihilation and one-loop 2-body radiative annihilation. Instead, for the operators of Eqs.~(\ref{Opsi1DMY}) and (\ref{Opsi1DML}) both features appear at same coupling and loop order, from the fact that they involve a scalar boson in the final state or its vacuum expectation value.}

Second, they bring a ``neutrino IB'' contribution to the neutrino energy spectrum,\footnote{For other setups with neutrinos from three-body decay, see \cite{3bdyOpp}.} parameterized by $\Gamma_{\nu}^{3b}$, which is 
not as peaked, as it scales as 
\be
\frac{dN}{dE_\nu} = 
\frac{32}{m_{DM}} \left(\!1\!-\!\frac{2E_\nu}{3 m_{DM}}\!\right) \left( \frac{E_\nu}{m_{DM}} \right)^2  \Theta\left(\dfrac{m_{DM}}{2}-E_\nu\right).
\label{IBnuspectrum}
\ee
However, it still displays a rise and a kinematical cutoff at $E_\nu=m_{DM}/2$ that are sharp enough to be clearly distinguished from expected astrophysical backgrounds of neutrinos.

Third, they bring an additional source of lower energy continuum of CRs (photons, antiprotons, positrons and neutrinos) from the $Z, W$ and $h$ decays together  with the leptonic final states.
The $Z/W$ and the lepton spectra are as in Eq.~\eqref{IBgspectrum} and \eqref{IBnuspectrum}, respectively (up to neglected corrections from the mass of these particles). For the CR spectra calculations, the scalar  (including longitudinal $Z_L$ and $W_L$) spectrum is also relevant, and is given by
\be
\frac{dN}{dE_\phi} = 
\frac{32 E_\phi}{m_{DM}^2} \left(1  - \frac{3 E_\phi}{m_{DM}}  +   \frac{2 E_\phi^2}{m_{DM}^2}  \!\right)   \Theta\left(\dfrac{m_{DM}}{2}-E_\phi\right).
\label{Phispectrum}
\ee

Bounds can be obtained separately on these three contributions --- the line-like photon spectrum, the line-like neutrino spectrum and the lower energy continuum of CRs --- from searches of $\gamma$-lines, $\nu$-lines  and CR-continuum signals, respectively.  Then, from an operator's given BRs, each such bound can be converted to a bound on any of the other partial decay widths.

Note that we call a particle ``primary" if  produced directly from one of the local effective operators, while if produced subsequently we call it ``secondary''.

\subsection{Photons' sharp spectral features} \label{subsec:photon sharp}
\begin{figure}[t]
\center{\includegraphics[width=0.97 \columnwidth]{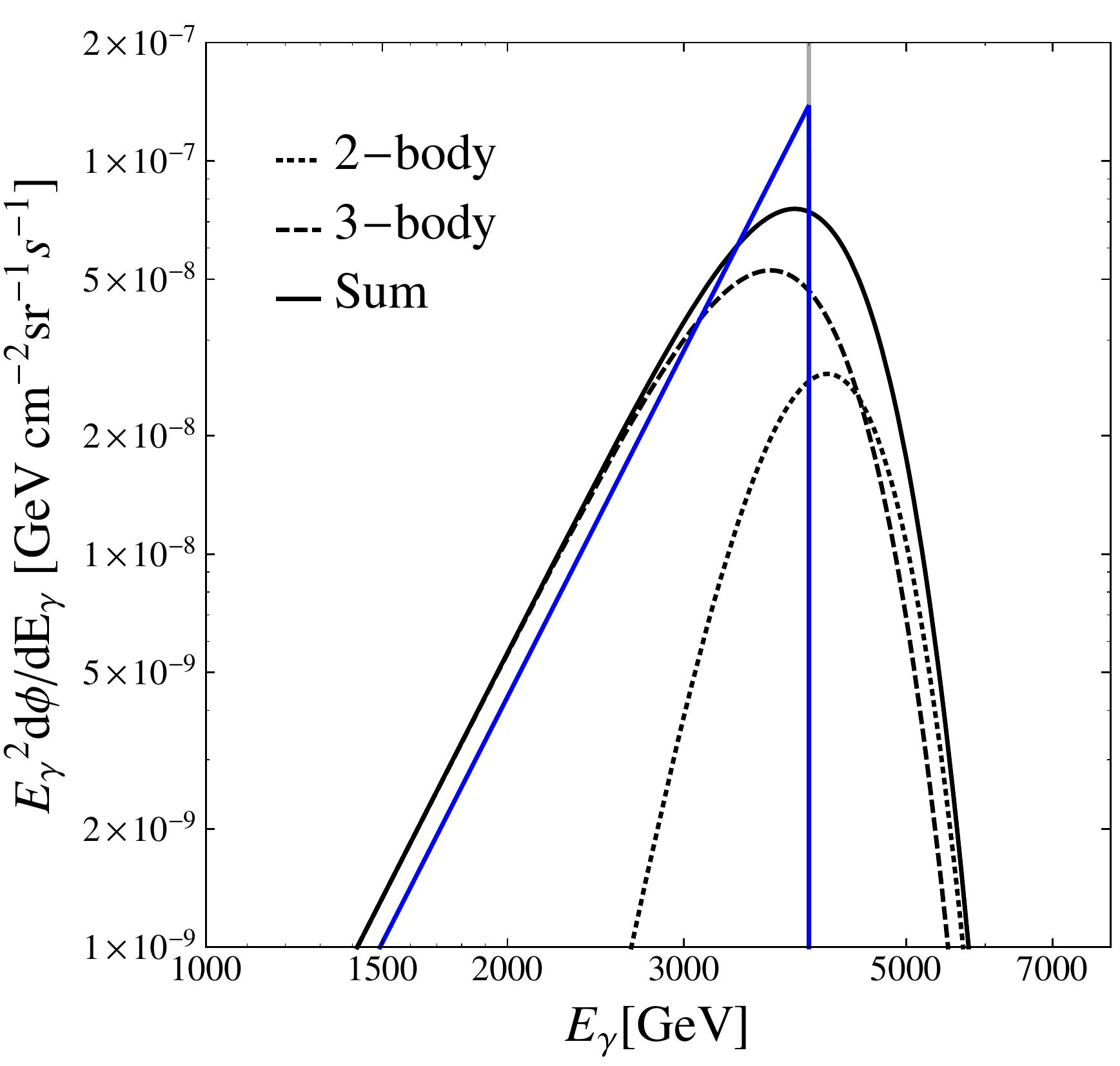}}
\caption{Primary photon energy spectrum obtained for the operator $O^{1Y}_{H,1/3-\text{let}}$ in Eq.~(\ref{A3body}) for an 8 TeV DM particle with a lifetime of $10^{28}$~s. The 2-body $\gamma\nu$ channel line (grey) and 3-body IB (blue) contributions are shown. The corresponding smeared spectra by a gaussian energy resolution of $15\%$, according to Eq.~(\ref{eq:Kernel}), as well as
 their sum are also shown (dotted, dashed and black curve, respectively). }
\label{fig:IBspectrum}
\end{figure}

In Fig.~\ref{fig:IBspectrum}, we show the characteristic primary photon energy spectrum induced at the 3-body decay level. The example shown is for the operator of Eq.~(\ref{A3body}) with $m_{DM}= 8$~TeV and total decay width $\Gamma_\text{tot} = 10^{-28} ~\text{s}^{-1}$.  
A proper determination of the constraints which hold on such a spectrum would require a dedicated analysis directly performed from data.
In the absence of such an analysis, we can nevertheless derive approximate bounds based on the following. If a telescope has a poor energy resolution it can initially not distinguish this line-like 3-body contribution from a monochromatic signal.  In this case, most of the primary 3-body decay photons have energies around $E_\gamma=m_{DM}/2$ within an energy range not wider than the energy resolution (together  with the monochromatic photons  from the 2-body contribution). In the opposite limit of very good energy resolution, the $\gamma$-line and the bulk of the primary photons from the  3-body decay are resolved to be spread at different energies.  In practice, with the current experimental energy resolutions, $r_E\sim10\%$ for Fermi-LAT \cite{Ackermann:2015lka} and $r_E\sim15\%$ for H.E.S.S. \cite{Abramowski:2013ax}, one is in an intermediate situation. In order to further quantify this, we need to know what are the respective contribution to the number of hard photons expected within the energy bin around $m_{DM}/2$.
Within a bin defined by $E_{min}=\frac{m_{DM}}{2}(1- r_E)$ and $E_{max}=\frac{m_{DM}}{2}(1+ r_E)$, this ratio is given by
the quantity $f_\gamma$ defined as follows 
\begin{equation}
f_\gamma=\frac{\displaystyle\int_\mathrm{bin}\!\! dE'  \int \!dE \; \frac{dN_\gamma^{3b}}{dE} \,K(E',E) } 
                         {\displaystyle\int_\mathrm{bin}\!\! dE' \int\! dE \; \frac{dN_\gamma ^{2b}}{dE} \,K(E', E)}.
\label{eq:fratio}
\end{equation}
with $dN_\gamma^{ib}/dE$ the gamma-ray spectrum of the  primary photons produced in an {\it i}-body decay, and $E'$ the reconstructed energy. For the detector response, we assume a gaussian function
\be
K (E',E) = \dfrac{1}{r_{E} E\sqrt{2 \pi}} e^{-\frac{1}{2}\left(\!\frac{E-E'}{r_E E}\!\right)^2}.
\label{eq:Kernel}
\ee
With this setup, our numerator depends on the detector resolution, while the denominator of Eq.~(\ref{eq:fratio}) stays fixed to  68\%. For $r_E=10$\% ($r_E=15$\%), we get $f_\gamma=0.43 (0.57)$. 
\begin{figure}[t]
\center{\includegraphics[width=0.95 \columnwidth]{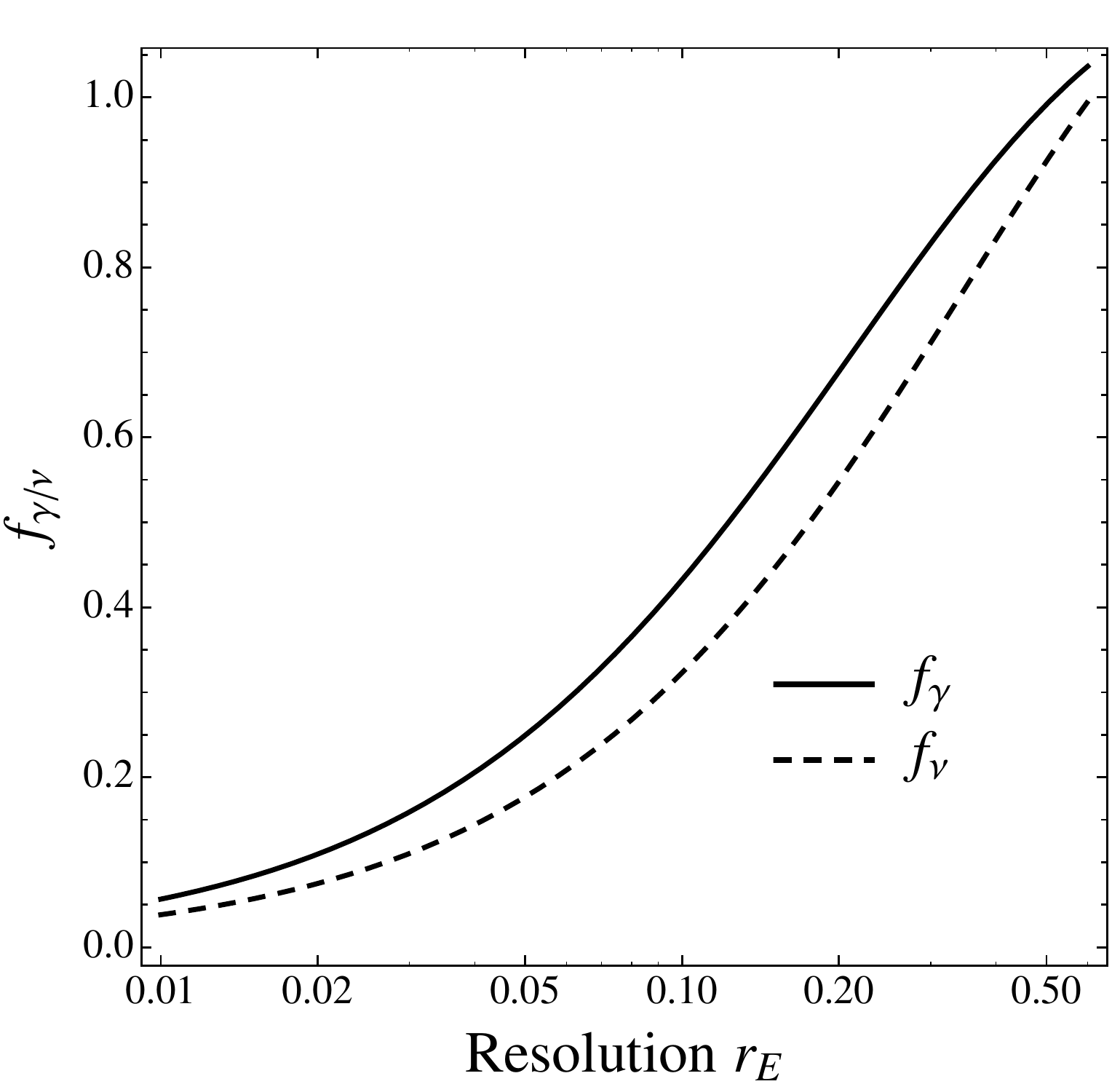}}
\caption{\textit{Solid line} ($f_\gamma$): the ratio between the number of prompt photons in 3-body decays and prompt photons in 2-body decays in an energy window $(m_{DM}/2)(1\pm r_E)$. \textit{Dashed line} ($f_\nu$): same ratio, but for neutrinos. The $f_{\gamma,\nu}$ are defined in Eq.~(\ref{eq:fratio}) and are only functions of the energy resolution $r_E$. The 3-body $\gamma$ and $\nu$ energy spectra are given in Eq.~\eqref{IBgspectrum} and \eqref{IBnuspectrum}, respectively. }
\label{fvsRes}
\end{figure}
In Fig.~\ref{fvsRes}, we plot $f_\gamma$ as a function of $r_E$.

\begin{figure}[t]
\vspace{-0.5cm}
\center{\includegraphics[width=0.98 \columnwidth]{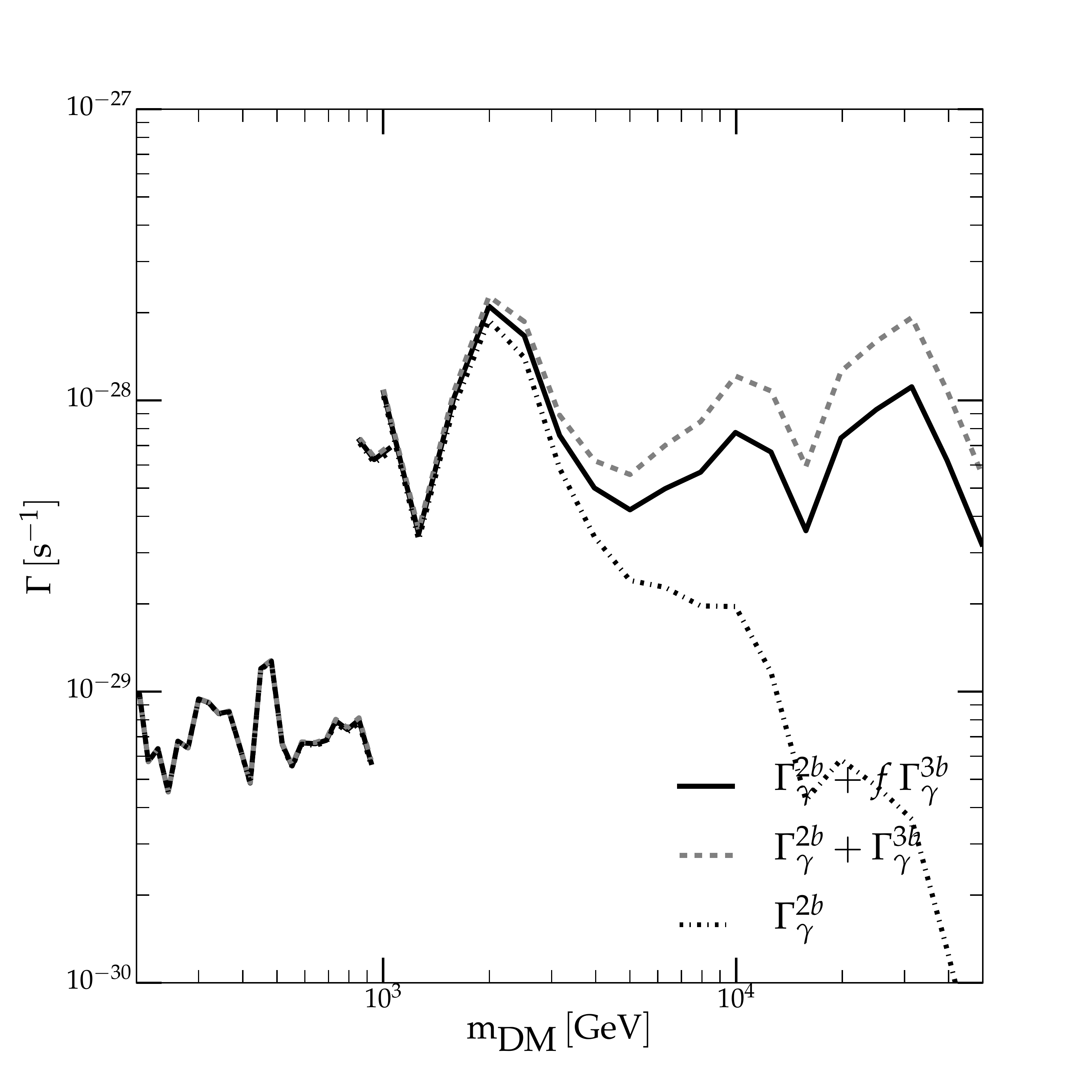}}
\caption{Solid black lines: 95$\%$ CL limits on the intensity of a $\gamma$-line from Fermi-LAT  (below $\sim$1~TeV) and H.E.S.S.~(above $\sim$1~TeV). These bounds are now reinterpreted as bounds on $\Gamma_\gamma^{2b}+f_\gamma \cdot \Gamma_{\gamma}^{3b}$ with $f_\gamma \simeq 0.43$ for Fermi-LAT  and $f_\gamma \simeq 0.57$ for H.E.S.S.. 
As the ratios of the various partial decay widths are totally fixed by the value of $m_{DM}$ for a given operator, these bounds can be translated into bounds on other partial widths. As an example we show the bounds induced on the 2-body decay width $\Gamma_\gamma^{2b}=\Gamma_{\gamma\nu}$ and on the total radiative decay width $\Gamma_\gamma=\Gamma_{\gamma}^{2b}+\Gamma_{\gamma}^{3b}$, for the operators of Eqs.~(\ref{Opsi1DMY}) and~(\ref{Opsi1DML})  with $\phi$ the SM scalar doublet. These bounds turn out to be identical for all the cases considered in Tab.~\ref{tab:BR}. Instead, the bound  this gives on the total DM decay rate $\Gamma_\text{tot}$ depends on the operator, and we do not show it here.}
\label{IBhardgammabounds}
\end{figure}

\begin{figure}[t]
\center{\includegraphics[width=0.9 \columnwidth]{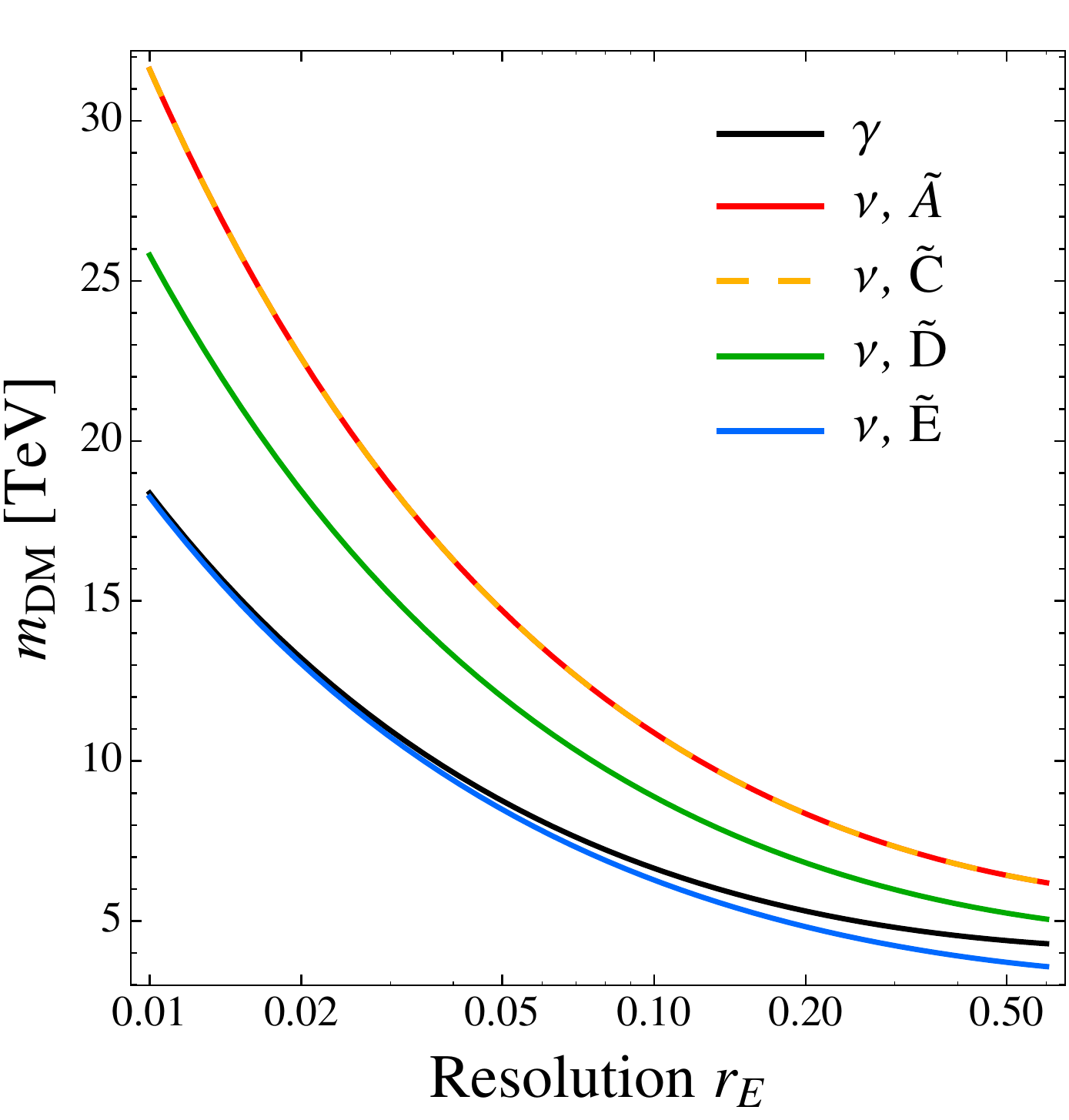}}
\caption{
Dark matter mass $m_{DM}$ above which, within the energy bin $(m_{DM}/2)(1\pm r_E)$, the integrated line-like photon/neutrino flux from 3-body decays gets larger than the monochromatic line from 2-body decays --- i.e., when $\Gamma^{3b}_{\gamma, \nu} / \Gamma^{2b}_{\gamma, \nu} \times f_{\gamma,\nu}$ becomes larger than $1$. For the photon spectra the result is the same for all operators (black curve), while for the neutrino spectra  we show the result (colored lines) for the $\tilde{A}$, $\tilde{C}$, $\tilde{D}$ and $\tilde{E}$ cases of Eqs.~(\ref{A3body})-(\ref{E3body}).}
\label{massvsRes}
\end{figure}

For both Fermi and HESS, the spectra from the 3-body final states can be considered as highly peaked in this energy bin. As a result, the observational bounds on a pure $\gamma$-line intensity can approximately be used for these cases too. However, the interpretation of these limits changes. Instead of being bounds on the 2-body $\gamma$-line decay width, $\Gamma_{\gamma}^{2b}$, these observational bounds now apply to $\Gamma_{\gamma}^{2b}+f_\gamma \cdot \Gamma_{\gamma}^{3b}$ with $f_\gamma$ as defined above. Given the fact that BRs are known, such limits can be translated into bounds on the other partial decay widths, as well as a bound on the total decay rate $\Gamma_\text{tot}$.  Fig.~\ref{IBhardgammabounds} shows the limits obtained on the line part $\Gamma_{\gamma}^{2b}$ and the total radiative decay width $\Gamma_{\gamma}^{2b}+\Gamma_{\gamma}^{3b}$. 
Given the fact that the 2-body BRs get tiny at large $m_{DM}$, without any surprise the bound it imposes 
on the pure line part, $\Gamma_{\gamma}^{2b}$, becomes extremely strong for the highest DM masses.
This implies that extremely good resolution would be required for a future experiment to resolve the pure line itself for high DM masses. For example, with a $r_E = 1\%$ energy resolution there are more IB photons than pure monochromatic line photons within an $(m_{DM}/2)(1\pm r_E)$ energy bin as soon as $m_{DM} \gtrsim 18$~TeV. In Fig.~\ref{massvsRes}, we give the value of this transition mass, as a function of $r_E$, when there are more primary photons from 3-body decay than monochromatic photons from 2-body final states in this highest energy bin.

\subsection{Neutrinos' sharp spectral features}
\begin{figure}[t]
\center{\includegraphics[width=0.95 \columnwidth]{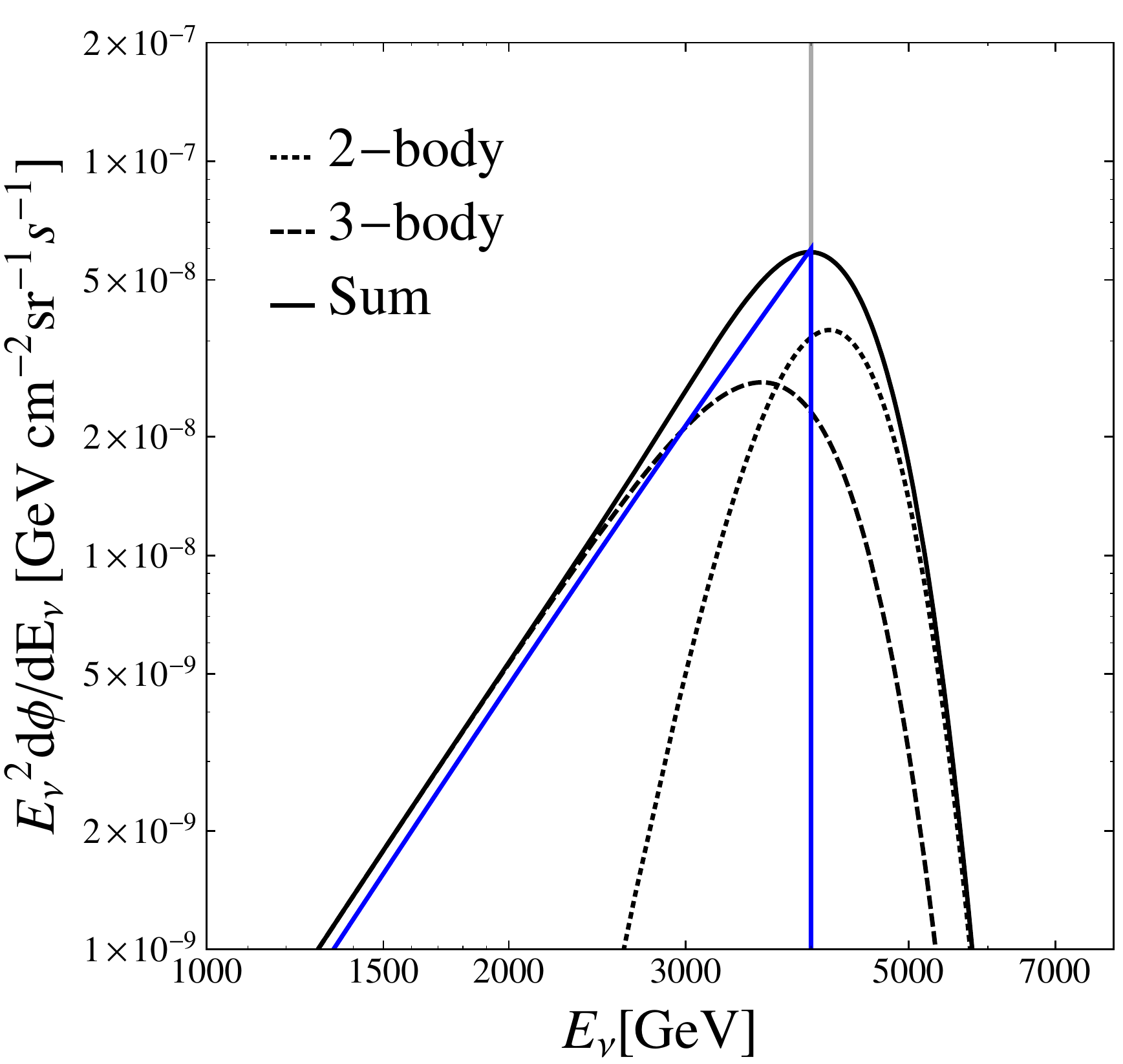}}
\caption{Primary neutrino energy spectrum obtained for the operator of Eq.~(\ref{Opsi1DMY}) for an 8 TeV SM singlet or triplet DM particle, 2-body $\Gamma_\nu^{2b}$ neutrino line contribution (black) and 3-body $\Gamma_\nu^{3b}$ IB contribution (blue). Corresponding two spectra to be observed by an experiment with $15\%$ energy resolution and sum of both contributions for this case. }
\label{fig:IBnuspectrum}
\end{figure}
Figure~\ref{fig:IBnuspectrum} shows the neutrino energy spectrum we get at 3-body decay level for a DM mass of 8 TeV and a 15$\%$ energy resolution. As can be seen by comparing it with the photon spectrum of Fig.~\ref{fig:IBspectrum} (obtained with same energy resolution), this spectrum is basically as sharp as the photon one for energies above $m_{DM}/2$ because it undergoes the same kinematical cutoff. As Eq.~(\ref{IBnuspectrum}) shows, it  scales as $(3/2)E_\nu^2m_{DM}-E_\nu^3$ instead of $E_\gamma^3$, but this mixture of a quadratic and cubic power law is still to be considered as a sharp feature (compared to the roughly $E^{-3}$ to $E^{-2}$ expected neutrino background spectrum).
This plot shows that it can be interesting to perform dedicated searches of both lines and IB type signals (as done in, e.g.\ Ref. \cite{Aisati:2015vma,Abramowski:2013ax}).  Here, we will however proceed in the same way as we did for the photons, by considering the number of 3-body neutrinos to be expected within the $(m_{DM}/2)(1\pm r_E)$ energy bin. That is, we use Eq.~(\ref{eq:fratio}) replacing  $\gamma$ with $\nu$. Figure~\ref{fvsRes} also gives the value of this $f_\nu$ (dashed line) as a function of the energy resolution $r_E$. It shows that for a typical energy resolution of $r_E = 15\%$ for the IceCube telescope, a fraction $f_\nu=0.44$ of the IB spectrum lies within this energy bin.
In a similar way to photons, the 2-body decay width $\Gamma_\nu^{2b}$ becomes negligible with respect to the 3-body width $\Gamma_\nu^{3b}$ for high $m_{DM}$ masses. 
The mass above which the primary neutrino contribution from 3-body decays exceeds the line signal from 2-body decays is shown in Fig.~\ref{massvsRes}. Unlike for photons, its value is operator-dependent and the different colored lines are for various operator predictions. The monochromatic neutrino bounds of Fig.~\ref{fermiicecube} have now to be reinterpreted as bounds on $\Gamma_\nu^{2b}+f_\nu \Gamma_\nu^{3b}$.

\subsection{Secondary cosmic ray constraints}
\begin{figure}[t]
\vspace{-0.5cm}
\center{\includegraphics[width=0.96 \columnwidth]{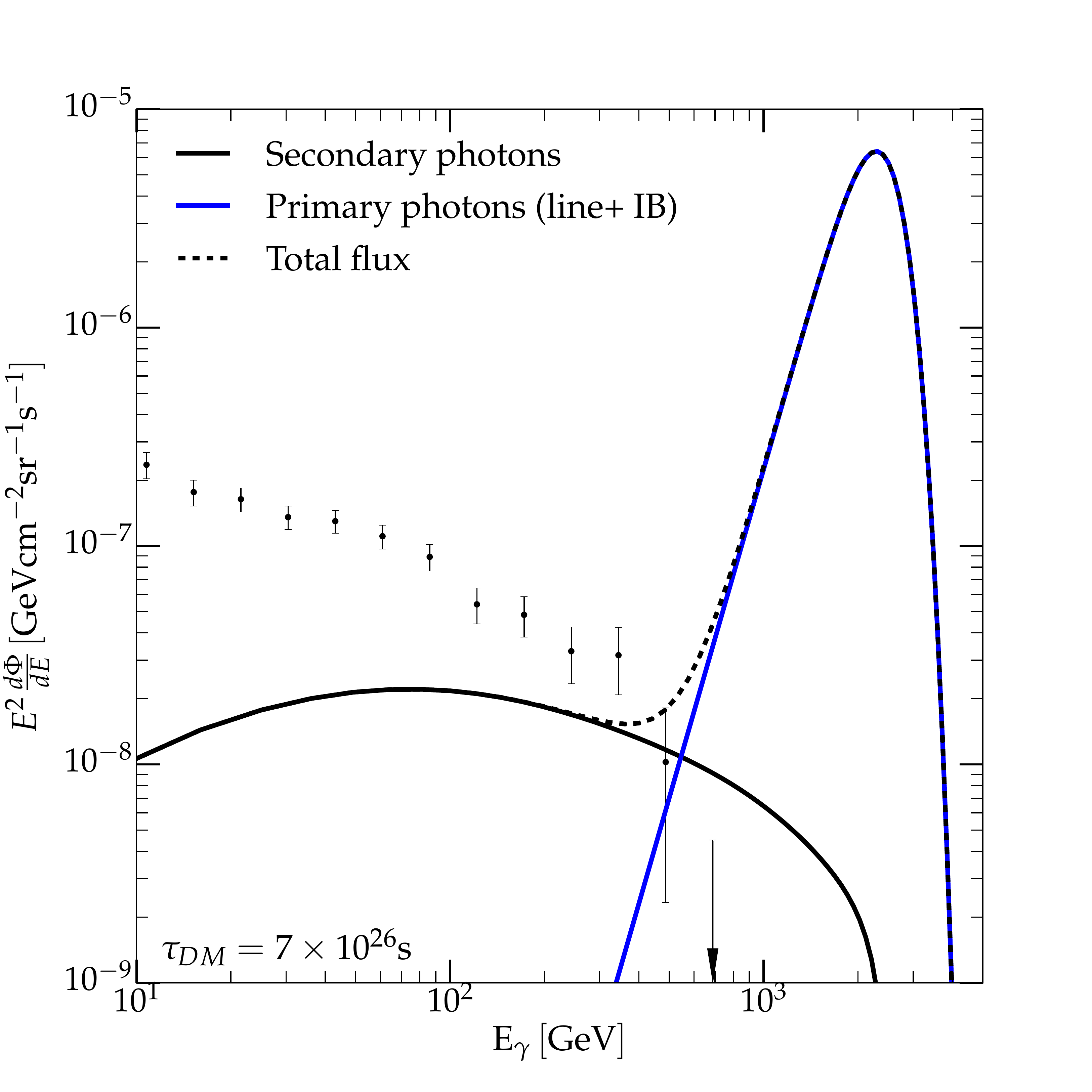}}
\caption{Photon flux obtained in the $\tilde{A}$ case of the operator of Eq.~(\ref{A3body}), for $m_{DM}=  5$~TeV, $r_E=15\%$ and a DM lifetime of $7 \times 10^{26}$ s. The line-like energy spectrum (blue), and the continuum spectrum from $Z$, $H$ and $W_L$ decay at lower energies (gray), together with the sum of these contributions (dashed), as constrained by the Fermi-LAT data.}
\label{IBspectrum2}
\end{figure}

As already said above, for operators involving the SM scalar field $H$, 3-body decay channels dominate over the 2-body ones if $m_{DM}$ is larger than $\sim 4$~TeV. 
As a result, these channels are expected to considerably increase the amount of low energy CRs, hence to considerably strengthen the associated bounds on monochromatic line signals from these operators.
Incorporating the CR contributions of all 2 and 3-body decay channels, Fig.~\ref{IBspectrum2} shows the corresponding secondary photon spectrum in the $\tilde{A}$ case of Eq.~\eqref{A3body} together with the line-like signal.
The derivation of the CR spectrum is as in Section~\ref{Sec:associatedCR}, with the difference that now primary particles from 3-body decays have a distribution in energy.

By imposing that the CR fluxes do not exceed the  isotropic gamma-ray background or antiproton measurements, we can derive upper bounds on each partial decay widths (given the fact that the ratios of the various partial decay widths are totally fixed for each operator for a given $m_{DM}$). In particular, the partial decay width to the line signal  can be constrained (as in Section~\ref{Sec:associatedCR}). 
\begin{figure*}
  \centering
  \subfigure[$F_Y$: $\tilde{A}$ case]{\includegraphics[width=0.49 \textwidth]{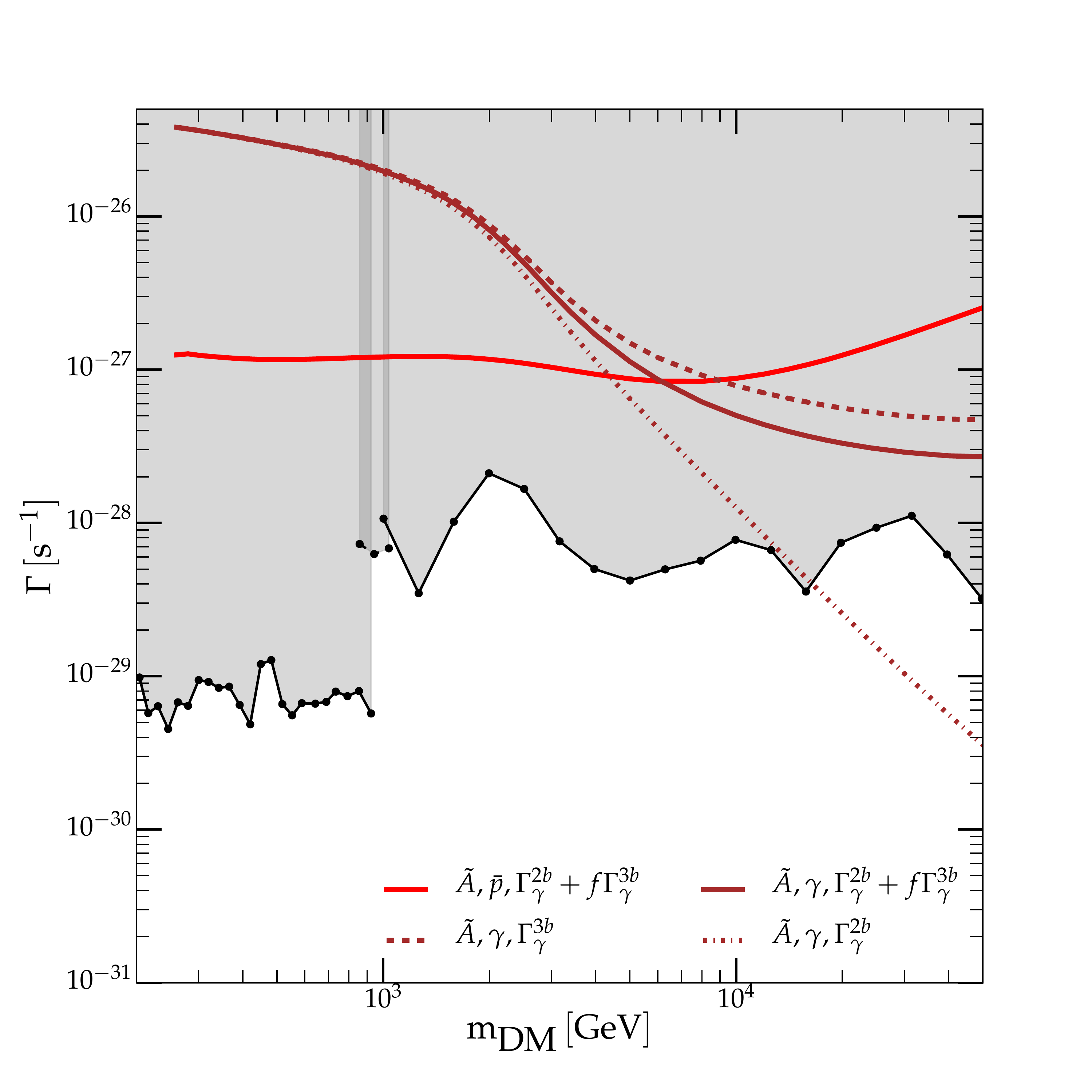}}\quad
  \subfigure[$F_L$: $\tilde{E}$ case]{\includegraphics[width=0.49 \textwidth]{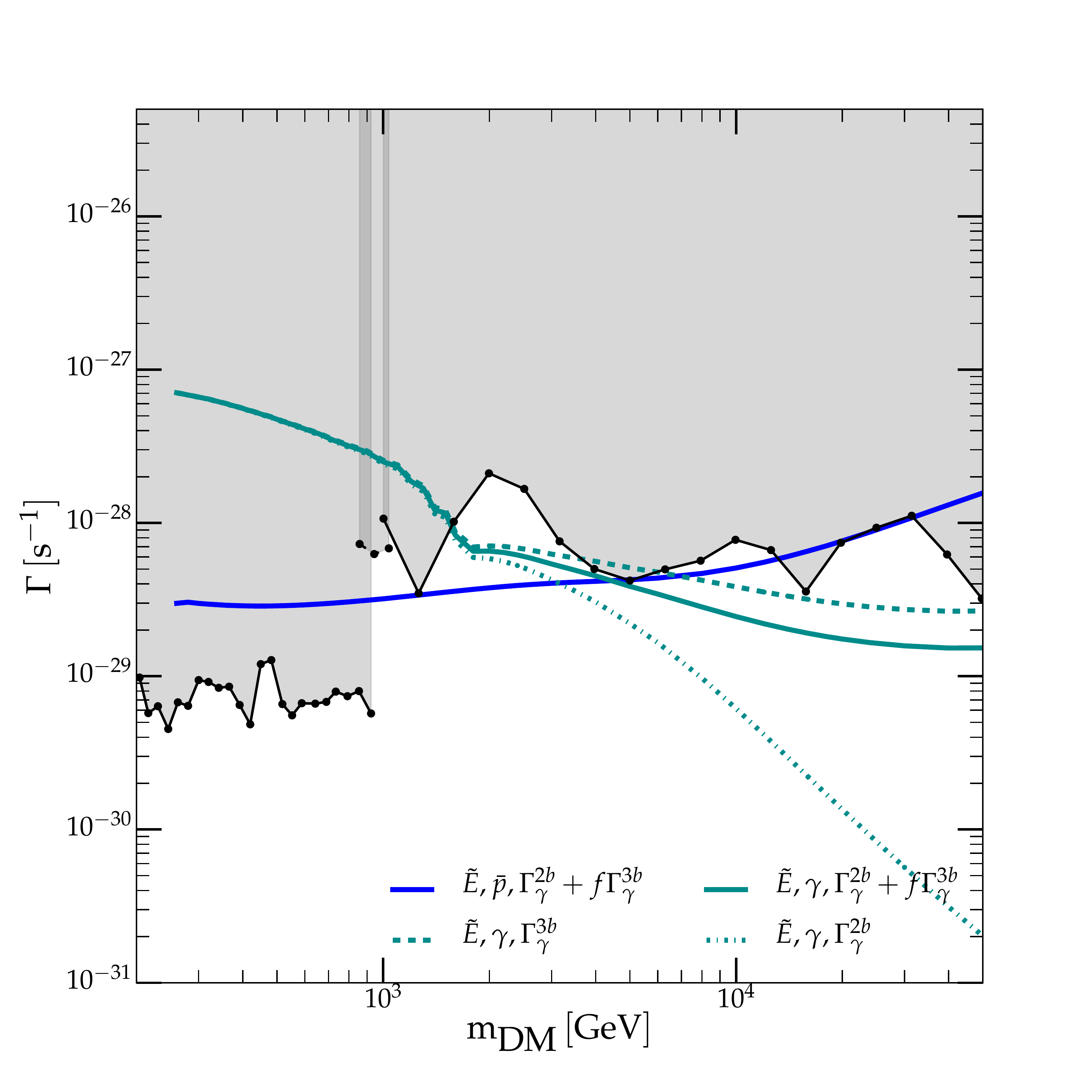}}
\caption{
95 \% CL limits on decay rates into a gamma-ray line-like signal. Direct line searches give the grey excluded regions, while the colored lines are the indirect limits on the line-like signal. \textit{Left panel:} Red solid curves present the limits on $\Gamma_{\gamma}^{2b}+f_\gamma \cdot \Gamma^{3b}_{\gamma}$ (with $f_\gamma=0.57$) derived by imposing that the associated CR prediction $\tilde{A}$ in Eq.~\eqref{A3body} does not overshoot PAMELA antiproton data (light red) or Fermi-LAT isotropic gamma-ray background data (solid dark-red). 
\textit{Right panel:}  The same  as the left panel (solid blue and solid cyan curve, respectively), but for the prediction $\tilde{E}$ of the operator in Eq.~(\ref{E3body}).  
We also give the limits on the decay width $\Gamma_{\gamma }^{2b} + \Gamma^{3b}_{\gamma }$ and on $\Gamma^{2b}_{\gamma}$ (using the BRs from Table~\ref{tab:BR} for these two operators).}
\label{CRIBbounds}
\end{figure*}
For operators leading to the $\tilde{A}$ and $\tilde{E}$ scenarios, Fig.~\ref{CRIBbounds} shows the  bounds implied by the continuum of gamma-rays,  on the $\gamma$ line (i.e.\ on $\Gamma_{\gamma}^{2b}$), on the decay width into a line-like signal ($\Gamma_{\gamma}^{2b}+\Gamma_{\gamma}^{3b}$) and its fraction that is actually be probed with pure monochromatic line-searches  ($\Gamma_{\gamma}^{2b}+f_\gamma \Gamma_{\gamma}^{3b}$; with $f_\gamma=0.57$). For the antiproton constraints we only show the bounds on $\Gamma_{\gamma}^{2b}+f_\gamma \Gamma_{\gamma}^{3b}$.
The bounds in this figure can be compared with the bounds in Fig.~\ref{CRIBbounds2body}, which was derived in the previous section considering only the 2-body decays.

One observes, as expected, that for high values of $m_{DM}$, the 3-body decay level bounds are more stringent than the bounds obtained at 2-body decay level. 
This is especially the case for the bounds one gets at 3-body decay level on the line part, $\Gamma_{\gamma\nu}$,
due to the $\sim m^2_{DM}/(64 \pi^2v^2_\phi)$ relative ratio discussed above.
This is less the case for the bounds on the total hard photon production $\Gamma_{\gamma}^{2b}+\Gamma_{\gamma}^{3b}$ or on 
$\Gamma_\gamma^{2b}+f_\gamma \cdot \Gamma_\gamma^{3b}$ to which this relative factor does not apply.
For the $\tilde{E}$ case of Eq.~(\ref{E3body})  and for $m_{DM}\gtrsim 1$~TeV, the implicit bounds on the line-like signal from lower energy photons are always stronger than the dedicated line search by H.E.S.S.
As Fig.~\ref{CRIBbounds} also shows, for the $\tilde{A}$ case of Eq.~(\ref{A3body}) the CR bounds are comparable to the direct line limits only if $m_{DM}\gtrsim 30$~TeV.
In other words, if one expects that telescopes will improve sensitivities by, say, one order of magnitude in the near future, the prospects of observing soon the pure $\gamma$-line part for these operators is low as soon as $m_{DM}\gtrsim 10$~TeV. 
But they could instead observe the characteristic IB spectral shape feature up to $E_\gamma\sim 50$~TeV, at the least.

\subsection{Comment on Line-like Signals from Radiative Electroweak Corrections }
\label{sec:EWline}
Similarly to our 3-body effects,  electroweak corrections by radiation off gauge bosons, as well as the decays of e.g.\  $Z$-bosons directly into neutrinos, can induce line-like features at the high energy end of the spectra (both in gamma-rays and neutrinos).  We studied these effects by relying on the electroweak corrections as implemented in ``PPPC'' and computed in Ref.~\cite{Cirelli:2010xx}. Regarding photon line-like features, the most sizable electroweak corrections  arise from the transversally polarized $W$-bosons with high energy in the final state. Therefore operators with F and $\tilde{E}$ predictions, which have the highest BRs to $W_T$ bosons, are the most affected among 2-body and 3-body final state decays, respectively. 

For the F prediction, these EW corrections increase the line-like signal (in a surrounding 15\% energy window) by up to a factor of 2 for the highest DM mass of 50 TeV (the BR into a pure gamma-line is 4.2\,\%  and the BR into $Wl$ is  80.2\,\%). For $m_{DM}$ below 5 TeV the effect is however less than $\sim$30\;\% on the line intensity.  The electroweak corrections also have some impact on our CR limits. For example, the small dips in the continuum CR limits at $m_\text{DM}\simeq 2$~TeV for the E and F cases in Fig.~\ref{CRIBbounds2body} is due to the extra hardening of the spectra from EW corrections and that these peak-like features coincide with the simultaneous drop in the measured isotropic gamma ray intensity around  1~TeV. 

For the  3-body predictions the line-like signal is not as peaked, and the continuum spectrum including electroweak corrections never contributes to the line-like signal by more than 30\,\% even for the $\tilde{E}$ case and our highest considered DM mass.

Similarly the line-like feature in neutrinos can be enhanced, but the effect is not as large (new narrow line-like structures can in principle appear from the charged leptonic final states) and increases the line intensity by at most $\sim$10\,\%. 

We  conclude that these electroweak corrections can be relevant in a more detailed study, and we included them for our CR continuum predictions. However they do not significantly alter the final line-like signals (at most a factor two) and we did not included  them for our prospects of double monochromatic (line-like)  signal strengths.

\begin{figure*}
  \centering
  \subfigure[ \;Constraints on $\gamma$ line-like signals]{\includegraphics[width=0.49 \textwidth]{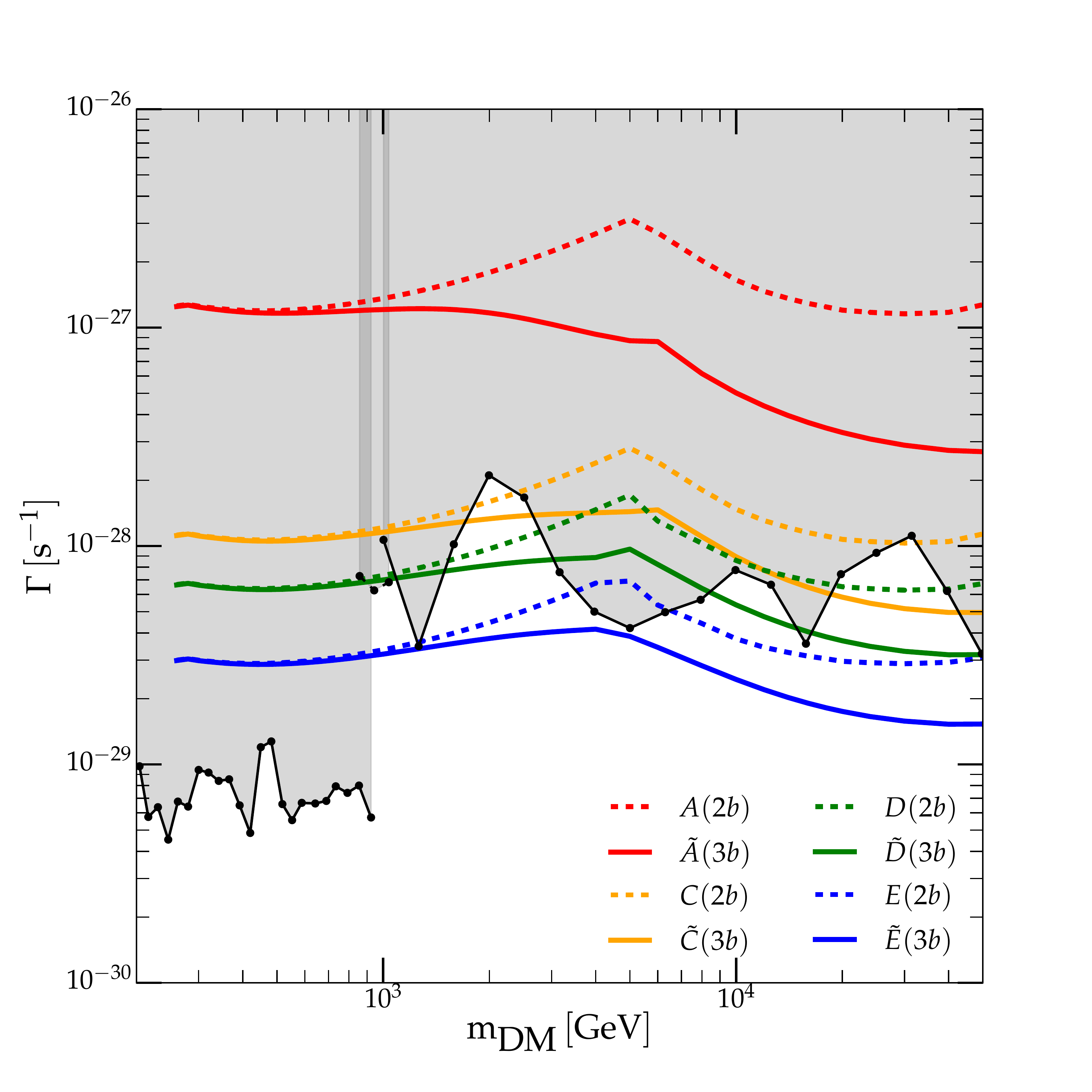}}\quad
  \subfigure[ \;Constraints on $\nu$ line-like signals]{\includegraphics[width=0.49 \textwidth]{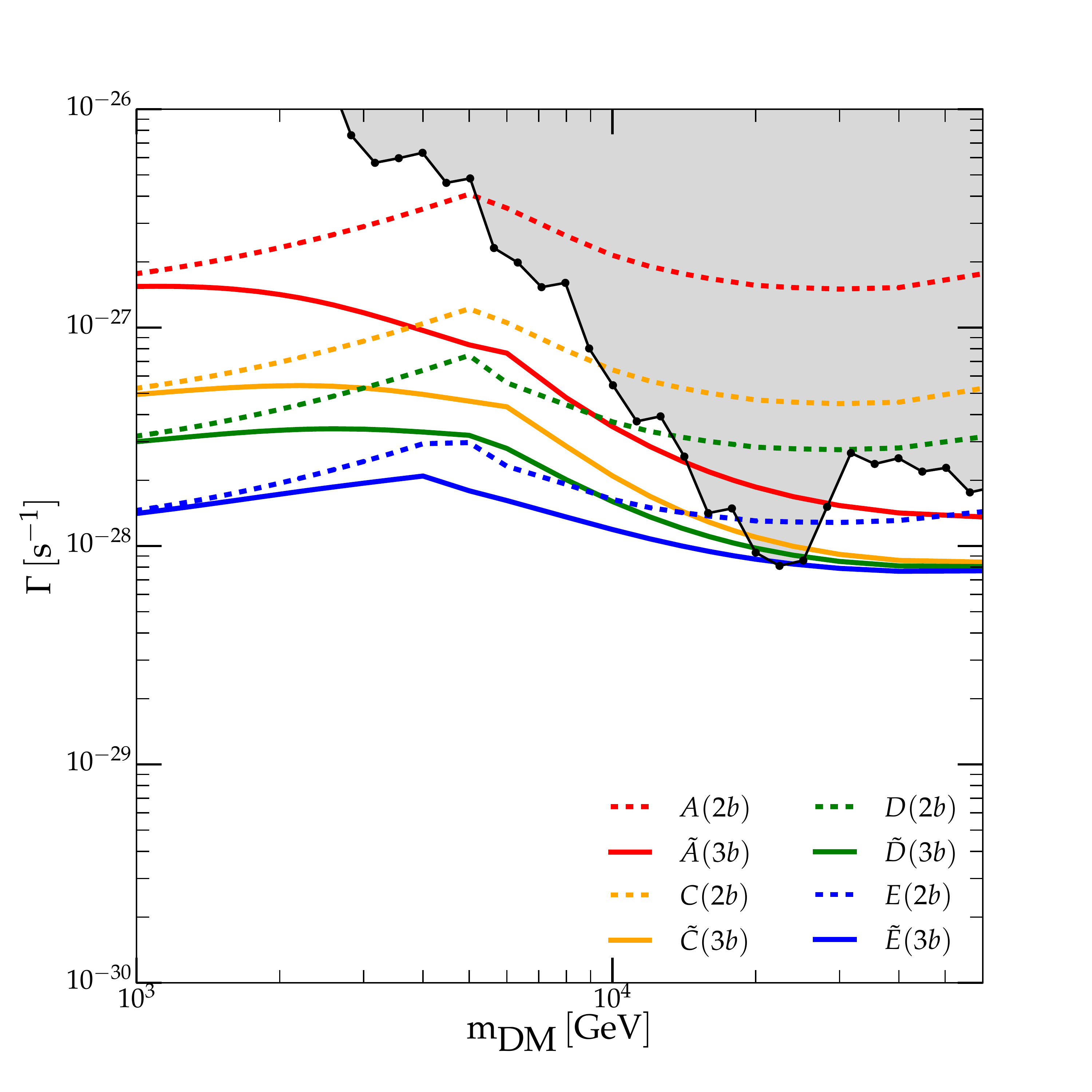}}
\caption{Summary of line-like signal limits from the various effective operators. \textit{Left panel:}  limits on gamma-ray line-like signals (as in Fig.~\ref{CRIBbounds2body} and \ref{CRIBbounds}). \textit{Right panel:} limits on neutrino line-like signals.  Dashed colored lines: the CR induced bounds on a line-like signal from operators with DM decays into 2-body final states. These operators and their predictions can be found in Tables~\ref{tab:opp}-\ref{tab:pheno}. Solid colored lines: same as dashed curves, but for the operators with relevant DM decays into  2- and 3-body final states. These colored solid curves are limits on $\Gamma_{\gamma}^{2b}+f_{\gamma,\nu} \cdot \Gamma^{3b}_{\gamma,\nu}$ for the CR predictions $\tilde{A}$, $\tilde{C}$, $\tilde{D}$ and $\tilde{E}$ (the operators of Eqs.~(\ref{A3body})-(\ref{E3body})) with $f_{\gamma}=0.57$ and $f_\nu=0.44$ from an assumed $15\%$ energy resolution in H.E.S.S.\ and IceCube, respectively.
These various limits should then be compared to the corresponding direct gamma- and neutrino line-search limits, shown by the grey exclusion regions.} 
\label{CRIBboundssummary}
\end{figure*}

\section{Double smoking gun prospects}
\label{Sec:prospects}

With the results obtained above, we can discuss the prospects of observing both a neutrino and gamma-ray line-like feature within the $DM\rightarrow \gamma \nu$ decay scenario. Let us proceed in three steps:

\medskip
I) In Fig.~\ref{CRIBboundssummary}.a, we summarize on a same plot the current limits on spectral $\gamma$-line features. The excluded grey regions come from the direct line-searches by the Fermi \cite{Ackermann:2015lka} and H.E.S.S.\ \cite{Abramowski:2013ax,Gustafsson:2013gca} collaborations. The solid colored curves are instead the upper bounds on the effective operators due to the continuum of low-energy CRs they induce, i.e.\ the bounds on $\Gamma_\gamma^{2b}$ considering only 2-body decays (dashed curves) and on $\Gamma_{\gamma}^{2b}+f_\gamma \Gamma_{\gamma}^{3b}$ including the case when 3-body contributions are relevant (solid curves). As said above, the former limits are valid for all operators (Table \ref{tab:opp},\ref{tab:pheno} and Fig.~\ref{CRIBbounds2body}) except if  the operator involves a scalar field and $m_{DM} \gg v_\phi$. For the case of relevant 3-body decays,  we show the representative $\tilde{A}$, $\tilde{C}$, $\tilde{D}$ and $\tilde{E}$ examples of Eqs.~(\ref{A3body})--(\ref{E3body}).

\medskip
II) Clearly, the CR bounds of Fig.~\ref{CRIBboundssummary}.a can also be translated into bounds on the intensity of a $\nu$-line feature, i.e.~on $\Gamma_\nu^{2b}$ for 2-body decays and on $\Gamma_\nu^{2b}+f_\nu \Gamma_\nu^{3b}$ for operators where 3-body decay channels dominate. By using the various BRs given in Tab.~\ref{tab:BR}, this gives the constraints of Fig.~\ref{CRIBboundssummary}.b, which can be compared to the direct neutrino line-search constraint from Fig.~\ref{fermiicecube}, also shown here by the grey region. Similarly to photons, this figure indicates, for each operator, by how much the neutrino line search sensitivity has to improve, at least, in order to observe a neutrino flux in the $DM\rightarrow \gamma  \nu$ scenario, without overshooting CR constraints.

\begin{figure}[t]
\center{\includegraphics[width=0.95 \columnwidth]{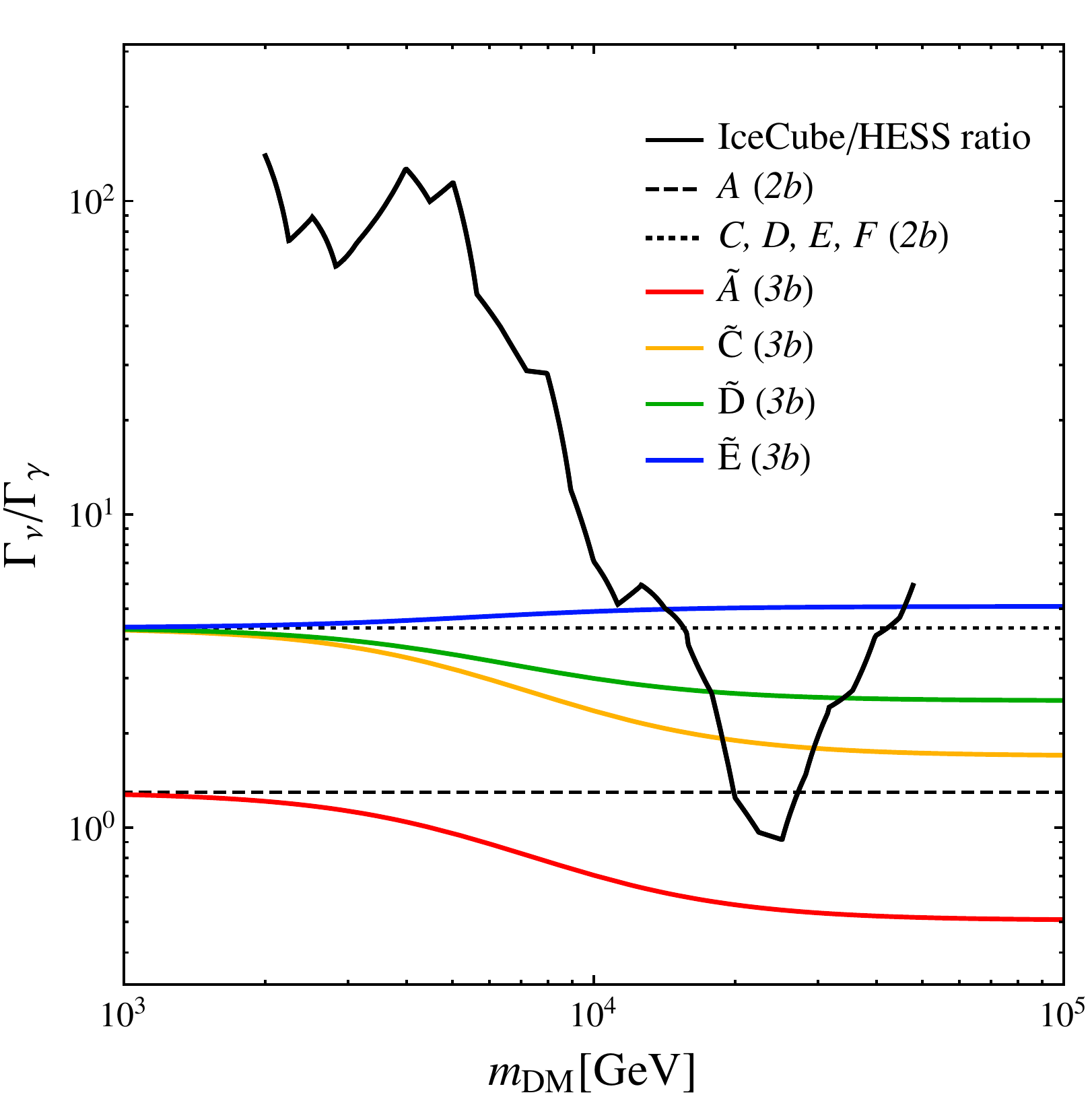}}
\caption{Summary of neutrino to photon sharp feature intensity ratios. Black line: ratio of neutrino to photon experimental sensitivities, from Fig.~\ref{fermiicecube}. For operators that do not involve a scalar field, this experimental ratio is to be compared with $\Gamma_\nu^{2b} / \Gamma_\gamma^{2b} =1.3 \text{~and~} 4.3$ (dashed and dotted line, respectively), obtained if the operator involves a $F_Y$ or a  $F_L$ field strength, respectively.  For operators that involve the SM scalar field $H$, this experimental ratio is to be compared with 
the $\Gamma_{\nu}^{2b}+f_\nu \Gamma_{\nu}^{3b}$ to $\Gamma_{\gamma}^{2b}+f_\gamma \Gamma_{\gamma}^{3b}$ ratio predicted by the various operators with $f$ calculated from 15\% energy resolution of the detectors (solid colored lines for $\tilde{A}$, $\tilde{C}$, $\tilde{D}$ and $\tilde{E}$ cases).}
\label{summaryratios}
\end{figure}

\medskip
III) From Fig.~\ref{CRIBboundssummary}.a and b, one observes that in the DM mass range from few~TeV to 50~TeV both the direct gamma- and neutrino-line searches constrain several  operators more than their associated CR fluxes do. For other operators it goes the other way around. One also notices that even if direct decay-rate limits are somewhat stronger on the  gamma-lines, the neutrino-line limits can still be as important to constrain an operator --- this is because many operators predict a relatively stronger neutrino than  gamma line-like signal (by a factor of one to five). The relative predicted neutrino to gamma-line strength, $\Gamma_\nu/\Gamma_\gamma$, is plotted in Fig.~\ref{summaryratios}. For operators which do not lead to relevant decays into 3 bodies, the ratio predictions are just 1.3 or 4.3, as given in Eqs.~(\ref{Ynugammaratioratio}) and (\ref{Lnugammaratioratio}). For operators which do involve relevant 3-body decays, we instead plot the predicted ratio of $\Gamma_\nu = \Gamma_{\nu}^{2b}+f_\nu \Gamma_{\nu}^{3b}$ to  $\Gamma_\gamma = \Gamma_{\gamma}^{2b}+f_\gamma \Gamma_{\gamma}^{3b}$ (solid colored lines), which, as said above, represents the line-like signal strength ratio. In the same plot we also show the ratio $\Gamma_\nu^\mathrm{limit}/\Gamma_\gamma^\mathrm{limit}$ of the neutrino to photon 95\%~CL line limits from Fig.~\ref{fermiicecube}. The factor between this observational (solid black) curve to an operator's prediction then indicates the minimal sensitivity improvement needed by IceCube relatively to H.E.S.S. to allow for simultaneous detections of a $\gamma$ and a $\nu$ line-like feature for that operator.

\bigskip
The exact prospects for detections of a double line depend on the operator considered. To quantify this in a few examples, it is useful to define $r_\gamma$ and $r_\nu$ as the ratios of the direct line-search limits ($\Gamma^\mathrm{limit}_{\nu,\gamma}$) to the indirect CR constraints on a line signal ($\Gamma^\mathrm{CR-limit}_\mathrm{\nu,\gamma}$) induced by each given effective operator setup:
\be
r_{\nu,\gamma} = \frac{\Gamma_{\nu,\gamma}^\mathrm{limit}}{\Gamma^\mathrm{CR-limit}_\mathrm{\nu,\gamma}}.
\ee
These ratios can directly be read off from Figs.~\ref{CRIBboundssummary}.a and \ref{CRIBboundssummary}.b.  The interpretation of these ratios is straightforward, as described below. 
If, for a given operator, the associated  CR constraint on the line signal is more stringent than those from direct searches of $\nu$ and $\gamma$ lines, then both $r_\nu$ and $r_\gamma$ are larger than 1. In this case, to detect both a $\nu$ and $\gamma$ line-like feature and stay compatible with {\it current} lower energy CR constraints, both sensitivities must be improved by factors larger than $r_\gamma$ and $r_\nu$, respectively.
This situation applies in particular to the $\tilde{E}$ case, except at the lowest DM masses where $r_\gamma < 1$.

If instead $r_\gamma$ is smaller than 1 and $r_\gamma<r_\nu$, then the $\gamma$-line feature could be just below present sensitivity, whereas a detection of the associated neutrino line feature would require that the neutrino sensitivities is improved by more than factor $r_\nu/r_\gamma$. Similarly, if $r_\nu$ is smaller than 1 and  $r_\nu<r_\gamma$, it means that the neutrino line could be just below current sensitivity and the gamma-line sensitivity would require an improvement by at least  a factor $r_\gamma/r_\nu$ to become observable for that operator prediction.
These ratios  $r_\nu/r_\gamma$ can also be read off from Fig.~\ref{summaryratios} by dividing the $\Gamma_\nu^\mathrm{limit}/\Gamma_\gamma^\mathrm{limit}$ (black solid line) by the operators'  predictions  $\Gamma_\nu/\Gamma_\gamma$ (colored, dashed and dotted lines).

\medskip
As an explicit example, the $\tilde{E}$ case (solid blue lines in Figs.~\ref{CRIBboundssummary}.a and \ref{CRIBboundssummary}.b) has both $r_\gamma$ and $r_\nu$ always larger than $\sim 1$ when $m_{DM}\gtrsim1$~TeV.
For example, with $m_{DM}= 10$~TeV, one has $r_\gamma=3$  and $r_\nu=5$. Hence the $\gamma$-line and $\nu$-line search sensitivities need to be improved by at least a factor $3$ and 5 respectively. A possible future observation of a photon or neutrino line with higher intensities could not stem from operators giving prediction $\tilde{E}$.

As another example, operators leading to the A case (i.e.\ 2-body, red dashed lines) could instead imminently give a double-line signal in gammas and neutrinos. For, say, $m_{DM}\simeq 30$~TeV it has $r_\gamma\sim r_\nu\sim 1/10<1$ and the associated CR signal must thus be at least 10 times smaller than what is probed today. For the same DM particle mass, $m_{DM} \simeq 30$~TeV, the C and D cases (i.e.~2-body, orange and green dashed lines)  instead have $r_\gamma\sim r_\nu\sim 1$ and an observation of both a double-line signal and an associated photon continuum flux could be just around the corner.

As for the 3-body $\tilde{A}$ case (red solid lines) with, say, $m_{DM}=3$~TeV, it gives $r_\gamma\simeq1/8$ and $r_\nu\simeq3$. This means that a photon line signal  can be  present just below current sensitivity, but the associated neutrino line-like signal would then still require to improve the $\nu$-line sensitivity by a factor $r_\nu/r_\gamma  \simeq 24$. 
This also means that the observation of a neutrino-line with a stronger intensity would rule out this setup.

\medskip
All in all, Figs.~\ref{CRIBboundssummary}.a, \ref{CRIBboundssummary}.b and \ref{summaryratios} show that improvements in sensitivities of the neutrino and  gamma-ray line searches by a factor of a ten allow for many possibilities of a {\it double smoking-gun evidence} of a DM particle without any tension with current lower energy CR constraints.

\section{The example of a minimal DM quintuplet}
\label{Sec:quintuplet}

As an explicit example, let us take a fermion quintuplet \cite{Cirelli:2005uq} with hypercharge $Y\!=\!0$. This ``minimal'' DM candidate is known to be accidentally stable --- as it couples linearly to SM fields only at the dimension 6 or higher level --- and to have its mass fixed by the relic density constraint to the value $m_{DM} = 9.6$~TeV \cite{Cirelli:2007xd}.  This scenario is in strong tension \cite{Cirelli:2015bda,Garcia-Cely:2015dda} with bounds on production of $\gamma$-line through DM annihilation, but still not totally excluded (depending on the DM halo profile considered). At the dimension to level, it can produce a $\gamma$-line through the single effective operator of Eq.~\eqref{D3body},
\begin{align}
&\text{$\tilde{D}$} :	  &&{\cal O}^{L1}_{H,\mathrm{5-let}} 	&&\hspace{-0.6cm}\equiv  &&\bar{L} \sigma_{\mu\nu} \psi_{DM}^\mathrm{5-let} F^{\mu \nu}_L H.  \nn  \,\,\,\,\,\,\,\,\,
\end{align}
The ratio of the pure monochromatic neutrinos to pure monochromatic photons, from 2-body decays, is therefore fixed in this framework to 
\be
R_{\nu/\gamma} \equiv\;\,  n_\nu/n_\gamma \simeq 4.3
\ee
and, as already mentioned above, at this level this operator leads to the $\tilde{D}$ case for the associated low energy continuum of CRs. Since it involves the SM scalar doublet, it has in addition a  3-body contribution which dominates for  $m_{DM}\gtrsim 4$~TeV. This implies both a photon IB  and a ``neutrino IB'' contribution. At three body decay level, the line-like signal ratio are therefore slightly below  4.3, as is shown by the $\tilde{D}$ line of Fig.~\ref{summaryratios} at $m_{DM} = 9.6$~TeV. The relative ratios of the nine DM decay channels we discussed are given in the quintuplet column of Tab.~\ref{tab:BR}.

The associated CR bounds which hold on line-like photon and neutrino features in this case are given in Figs.~\ref{CRIBboundssummary}.a and \ref{CRIBboundssummary}.b (solid green lines).
For $m_{DM}=9.6$~TeV, the H.E.S.S.\ line limits, which as explained above is interpreted to hold on $\Gamma_{\gamma}^{2b}+f_\gamma \cdot \Gamma_\gamma^{3b}$ with $f_\gamma \simeq 0.57$, is about a factor $r_\gamma\simeq 1$ weaker than the bound on the same quantity from the CR continuum constraint --- see Fig.~\ref{CRIBboundssummary}.a. For neutrinos the direct monochromatic line limit is about a factor $r_\nu\simeq 2$   weaker than the associated CR constraints, see Fig.~\ref{CRIBboundssummary}.b. These factors of improvements could be within reach in a not too far future \cite{Anchordoqui:2015lqa}.
If a neutrino or gamma line were to be observed just below current bounds, then the CR continuum signal should also be within reach or the signal could not be due to this quintuplet decay setup..

Also, note that for $m_{DM}=9.6$~TeV, the hard ``IB'' spectra intensities  within an $(m_{DM}/2)(1\pm r_E)$ energy bin dominates over the lines, unless $r_E$ is at least around $4\,\%$ for gamma-rays and 8\,\% for neutrinos.

\section{Millicharged DM option}

So far, we have implicitly assumed that the DM particle is electrically neutral. However, it is perfectly possible to give a small electric ``millicharge'' to the DM particle. Such a millicharge could arise from a kinetic mixing interaction between the $U(1)_Y$ field strength and a new $U(1)'$ field strength (whose gauge boson is a massless $\gamma'$) \cite{Holdom:1985ag} or from a Stueckelberg mixing mechanism \cite{Kors:2004dx} (whose associated gauge boson is a massive $Z'$). If the DM is millicharged, a new set of operators must be added to the list in Eqs.~(\ref{OpsiDM5Y}-\ref{Opsi3DML}). The extra effective operators we get are those with a covariant derivatives acting on fermion fields with a millicharge \cite{Aisati:2014nda}:
\begin{eqnarray}
&&D_\mu D_\nu\bar{L}\sigma^{\mu\nu}\psi_{DM}\\
&&D_\mu D_\nu\bar{L}\sigma^{\mu\nu}\psi_{DM} \phi\\
&&\bar{L}\sigma^{\mu\nu}D_\mu D_\nu\psi_{DM}\phi\label{eq:milipos}\\
&&D_\mu\bar{L}\sigma^{\mu\nu}D_\nu\psi_{DM}\phi
\end{eqnarray}

The phenomenology of these operators has been considered at length in Ref.~\cite{Aisati:2014nda}. If the fields to which  the covariant derivatives apply are not $SU(2)$ singlets, an observable $\gamma$-line is not feasible because the DM decay channel to $Z$ (and the $W$ channel, if any) is largely boosted with respect to the $\gamma$-line channel. The boost of the $Z$ (and $W$) channel is inversely proportional to the millicharge squared and therefore leads to too strong associated CR emissions. However, there is a single operator which escapes this constraint in the above list, namely 
\be
\text{$\tilde{A}$}: \quad\quad \bar{L}\sigma^{\mu\nu}D_\mu D_\nu\psi_{DM}\phi  \tag*{  (\ref{eq:milipos})},\quad\quad \nn
\ee 
with $\psi_{DM}$ a SM $SU(2)_L$ singlet and $\phi$, interestingly, having the same quantum numbers as the SM scalar doublet $H$.
As for the $\nu$-line signal instead, it can be generated by all the operators with a strong intensity, and thus observed, via the $\nu Z$ decay channel. 
For what concerns the neutrino to photon flux ratio, this unique operator gives in addition to DM decay into $\nu\gamma$ and $\nu Z$ also the decay into $\nu \gamma'$ for the kinetic mixing option and, if kinematically allowed, into $\nu Z'$ in the Stuckelberg case. These decays into the $U(1)'$ gauge boson and a neutrino, unlike the decay into $\nu\gamma$, are not suppressed by the value of the millicharge squared.
Given the bounds holding on millicharged DM, see e.g.\ Ref.~\cite{Aisati:2014nda}, this means that the intensity of the $\nu$ line, associated to the $\gamma'\nu$ channel, is orders of magnitude stronger than the $\gamma$ line for the kinetic mixing option (as well as for the Stuckelberg option if $m_{Z'}< m_{DM}$).
In other words,  for a millicharged DM candidate, it is possible to get similar monochromatic $\nu$ and $\gamma$ fluxes only for this unique operator with the two additional conditions that: i) the millicharge originates from a Stueckelberg mechanism and ii) that $m_{Z'}> m_{DM}$. In that case, since it is the hypercharge gauge boson that mixes with the $U(1)'$ gauge boson to give rise to a millicharge, the relative $\nu$- to $\gamma$-line intensity is 
\be
R_{\nu/\gamma} \;\equiv\; n_\nu/n_\gamma \simeq \; 1.3,
\ee
as in Eq.~(\ref{Ynugammaratioratio}). This case gives both $\gamma$ and $\nu$ lines and can have 3-body $\gamma$ and $\nu$ ``IB'' contributions. With $\phi=H$, it gives the same predictions of CR emission as the $\tilde{A}$ case, represented by the red solid curves in Fig.~\ref{CRIBbounds}.

\section{Summary}

From few TeV to 100 TeV energies, direct searches for monochromatic neutrinos from decaying DM are reaching a sensitivity comparable to those on monochromatic photons.  Motivated by this fact, we have studied the  possible predictions of a neutrino and photon line --- a double ``smoking-gun'' evidence for DM particles ---  within the particularly predictive scenario where both lines are emitted by the same process, namely $\psi_{DM}\rightarrow \nu \gamma$. 
In a systematic way, we considered the complete list of effective operators which lead to such decays. We found 10 operator structures which lead to 25 operators involving SM fields on top of the DM one (all listed in Table~\ref{tab:opp}).

Along this scenario, the neutrino to photon ratio is predicted to be within a factor of a few, or even totally predicted for cases where only one effective operator induces the decay.
The expected line shape in the photon and neutrino energy spectra depends on whether the operators involve a scalar field or not.
If the operator considered does not, the DM 2-body decays dominate,  and $\gamma$ and $\nu$-lines are truly monochromatic. If, instead, the operators involve a scalar field, for instance the SM scalar doublet, 3-body decay channels become dominant for $m_{DM}\gtrsim 4$~TeV. In these cases, the primary photon spectrum is not dominated anymore by the monochromatic $\gamma$-line at the highest energies,  but by ``IB'' contributions which also display a sharp feature in the spectrum.
Interestingly, these 3-body decays do not only give a photon IB spectrum but also a neutrino spectral peak. Given IceCube's performance improvements and good energy resolution for cascade like events, we stress that this kind of 3-body signal, within the framework considered, or other possible scenarios, could be put in evidence by the IceCube collaboration by accordingly search for such spectral features (as in Ref.~\cite{Aisati:2015vma}). Note that at the 2-body decay level we presented the phenomenology of all the 25 operator cases. At the 3-body decay level we considered 4 cases which are representative of what is the typical expected phenomenology. 

The prospects for a ``double smoking gun''  (or ``poly-monochromatic'') evidence of the DM particle have been discussed in Section \ref{Sec:prospects}. They can be read off from Figs.~\ref{CRIBboundssummary}.a, \ref{CRIBboundssummary}.b and \ref{summaryratios}, which summarize our results. Fig.\ref{CRIBboundssummary}.a and \ref{CRIBboundssummary}.b show the upper bounds which hold on the intensity of the $\gamma$ and $\nu$ sharp feature signals from secondary CR emission. These constraints must be compared with constraints from direct search for monochromatic signals, also shown in these figures. Figure~\ref{summaryratios} shows the ratio of $\nu$ to $\gamma$ line feature intensities predicted by the various operators. A comparison of these 3 plots (as explicitly done for a few examples in Section V) shows that there are clear possibilities for a ``double smoking gun'' discovery of the DM particle just around the corner.
If a gamma and neutrino line-like signal are observed, it would be interesting to  further explore the full range of linear combinations of the operators (listed in Table~\ref{tab:pheno} for each DM field).

\vfill
\begin{acknowledgments}
We thank L.~Covi, A.~Ibarra and M.~Tytgat for useful discussions. 
MG acknowledges partial support from the European Union FP7 ITN Invisibles (Marie Curie Actions, PITN-GA-2011-289442). TS thanks the HEP group at the Cavendish Laboratory for hospitality.
This work is supported by the FNRS-FRS, the FRIA, the IISN and the Belgian Science Policy, IAP VI-11.
\end{acknowledgments}

\appendix
\section{Comments on astrophysical constraints} \label{sec:A}
On top of the primary injected gamma-ray spectrum from DM decays, we also assessed the Inverse Compton contribution from CMB photons scattering off DM induced electrons and positrons  (see, e.g., \cite{Ando:2015qda}). This is however found to have a marginal effect (at most a factor 2, and then only for DM masses above 10 TeV and for the F case, so it was not included in our results of Figs. \ref{CRIBbounds2body}, \ref{CRIBbounds} and \ref{CRIBboundssummary}).
Recently, it has been  emphasized  that various astrophysical sources of photons, including blazars, can explain most of the continuum photon spectrum \cite{DiMauro:2015tfa}.
If true, this leads to stronger continuum photon constraints on DM decay.  Imposing that the DM induced flux cannot exceed the difference between the observed flux and the astrophysical contribution, as given in Fig.~1 of Ref.~\cite{DiMauro:2015tfa} (taking, to be conservative, the lower edge of the blue band given in the upper left panel of this figure) and only considering the bins where the DM signal overshoots the observed flux, one would improve the continuum photon bounds of Fig.~\ref{CRIBbounds2body} by about a factor 2. 

Our choice of the MED propagation setup \cite{Donato:2003xg}, relevant for the results of Figs. \ref{CRIBbounds2body}, \ref{CRIBbounds} and \ref{CRIBboundssummary}, is based on the recent preliminary AMS-02 data \cite{AMS22015,Giesen:2015ufa}, which seems to be more favored than the MIN setup. The MIN setup would give weaker and more  conservative constraints by a factor of about 5. Furthermore, the fact that the anti-proton background is assumed to exactly match the used PAMELA data \cite{Adriani:2010rc} in each energy bin (following \cite{Gustafsson:2013gca,Cirelli:2013hv} we considered data above 10 GeV) means that limits might be more aggressive than if larger background uncertainties are included.  With new AMS-02 data, we also reassessed our positron constraints (see, e.g., \cite{Ibarra:2013zia}), and concluded that they are always weaker than our anti-proton constraints because  positrons come together with $W$ bosons, which themselves produce antiprotons in quantity.

Regarding neutrinos, we have also ensured that the induced continuum of low energy neutrinos from EW corrections does not constrain our operators more than any other of our constraints.

 
\end{document}